\documentclass[a4paper,11pt]{article}
\pdfoutput=1 

\usepackage{jcappub} 
\DeclareUnicodeCharacter{2212}{-}
\usepackage[T1]{fontenc} 
\usepackage{graphicx}
\usepackage{subfigure}
\graphicspath{{img/}}
\usepackage{amsmath}
\usepackage{amssymb}
\usepackage{amsfonts}
\usepackage{mathrsfs}
\usepackage{lscape}
\usepackage{hyperref}
\usepackage{verbatim}
\usepackage{xcolor}
\usepackage{bm}
\usepackage{tabularx}
\usepackage{mathtools}
\usepackage{ulem}
\newcommand{\be}{\begin{equation}}
\newcommand{\ee}{\end{equation}}
\newcommand{\ba}{\begin{eqnarray}}
\newcommand{\ea}{\end{eqnarray}}
\newcommand{\nn}{\nonumber}

\newcommand{\Lc}{\mathcal{L}_{\mathrm{cdm}}}
\newcommand{\tildeLc}{\tilde{\mathcal{L}}_{\mathrm{cdm}}}
\newcommand{\trhoc}{\tilde{\rho}_{\mathrm{cdm}}}
\newcommand{\rhoDE}{\rho_{\mathrm{DE}}}
\def\bea{\begin{eqnarray}}
\def\eea{\end{eqnarray}}
\def\eqi{\begin{equation}}
\def\eqf{\end{equation}}
\def\eqia{\begin{eqnarray}}
\def\eqfa{\end{eqnarray}}
 
\def\Mpl2{M_{\rm{pl}}^2}
\def\cH{\mathcal{H}}

\definecolor{darkgreen}{rgb}{0,0.6,0}

\newcommand{\mc}{\mathcal}
\renewcommand{\(}{\left(}
\renewcommand{\)}{\right)}
\renewcommand{\[}{\left[}
\renewcommand{\]}{\right]}

\makeatother

\title{Scrutinizing coupled vector dark energy in light of data}

\author[\star, 1]{Wilmar Cardona,\note{Corresponding author.}}
\author[\dagger, 2]{Jose L. Palacios-C\'ordoba}
\author[\mathparagraph, 2]{C\'esar A. Valenzuela-Toledo}

\affiliation[1]{Departamento de F\'isica, Universidad Antonio Nari\~no, Cra $3$ Este No. $47$A-$15$, Bogot\'a DC, Colombia}

\affiliation[2]{Departamento de F\'isica, Universidad del Valle, Ciudad Universitaria Mel\'endez, 760032, Cali, Colombia}

\emailAdd{$\star$ wilmarcardonac@gmail.com}
\emailAdd{\mbox{$\dagger$ palacios.jose@correounivalle.edu.co}}
\emailAdd{\mbox{$\mathparagraph$ cesar.valenzuela@correounivalle.edu.co}}

\date{\today}

\abstract{
Since current challenges faced by $\Lambda$CDM might be hinting at new unravelled physics, here we investigate a plausible cosmological model where a vector field acts as source of dark energy. In particular, we examine whether an energy-momentum exchange between dark energy and dark matter could provide an explanation for current discrepancies in cosmological parameters. We carefully work out equations governing background and linear order perturbations and implement them in a Boltzmann code. We found that a negative coupling makes the dark energy equation of state less negative and closer to a cosmological constant during the matter dominated epoch than an uncoupled vector dark energy model. While the effect of the coupling is hardly noticeable through its effect on matter density perturbations, matter velocity perturbations and gravitational potentials are enhanced at late-times when dark energy dominates. Therefore, data of redshift space distortions help to narrow down these kinds of couplings in the dark sector. We computed cosmological constraints and found common parameters also present in $\Lambda$CDM are in good agreement with the Planck Collaboration baseline result. Our best fit for a negatively coupled vector field predicts a higher growth rate of matter perturbations at low redshift, thus exacerbating the disagreement with redshift space distortions data. While a positively coupled vector field can lead to power suppression of $P_{\rm{m}}(k,z=0)$ on small scales as well as a lower growth rate of matter perturbations than the standard model, it might compromise the goodness of fit to the CMB angular power spectrum on small scales. We conclude that our negatively coupled vector dark energy model does not solve current tensions (i.e., $H_0$ and $\sigma_8$). Moreover, having three additional parameters with respect to $\Lambda$CDM, the negatively coupled vector dark energy model is heavily disfavoured by Bayesian evidence.   
}
\makeatletter
\gdef\@fpheader{}
\makeatother
\begin{document}
\maketitle
\flushbottom

\section{Introduction}
\label{Sec: Introduction}

We may not be fully convinced of its success, but the current standard model of cosmology $\Lambda$CDM is the best description of the Universe we have at our disposal \cite{Planck:2018vyg,PhysRevLett.122.171301,Sawala:2022xom}. The concordance model not only is in good agreement with most astrophysical observations, due to a reduced number of degrees of freedom it also performs much better than its rivals when considering a Bayesian model comparison \cite{Heavens:2017hkr,Dutta:2019pio}. The price to pay for its undeniable success is high since $\Lambda$CDM is just a phenomenological model challenging our knowledge of fundamental physics condensed on the Standard Model of particle physics \cite{Weinberg:1988cp,Carroll:2000fy}.     

Despite of being a great achievement, many physicists are not fully satisfied with the explanations provided by the standard cosmological model. Firstly, it relies on the existence of the so called  Cold Dark Matter (CDM), which thus far remains directly undetected on experiments carried out on Earth \cite{Bertone:2016nfn}. Although known dark matter candidates such as neutrinos or black holes exist, they hardly account for the amount of dark matter required to explain current observations ($\approx 30\%$ of the total energy budget). Secondly, the currently most abundant energy ingredient in the concordance model, dubbed Dark Energy (DE) and incarnated as a cosmological constant $\Lambda$, is a major problem. Predictions for $\Lambda$ from Quantum Field Theory heavily mismatch the observational constraint, a conundrum for fundamental physics~\cite{Martin:2012bt}.     

Thus far the lack of conclusive evidence makes the nature of both CDM and DE pretty much unclear. Although in the literature it is possible to find Relativistic Modified Newtonian Dynamics (RMOND) models as alternatives to CDM \cite{PhysRevLett.127.161302}, most CDM particle candidates emerge in physics Beyond the Standard Model~\cite{Bertone:2004pz}. As for DE, the main alternatives to the cosmological constant are dynamical fields~\cite{Copeland:2006wr} and modifications to General Relativity (GR)~\cite{Clifton:2011jh,Bamba:2012cp}. Besides the puzzling nature of both CDM and DE, an increasing experimental sensitivity has started to reveal pretty interesting discrepancies in cosmological parameters of the standard cosmological model. Nowadays perhaps the most intriguing tension involves the expansion rate of the Universe $H_0$ which differs by $\approx 5\sigma$ as measured by low- and high-redshift probes. This difference could be pointing towards new physics disregarded in the $\Lambda$CDM model~\cite{Riess:2019cxk,Freedman:2021ahq,Riess:2021jrx}. A similar more moderate discordance regards the strength of matter clustering $\sigma_8$ which could also be a hint for a more detailed physical modelling~\cite{10.1093/mnras/stac2429,Asghari:2019qld,Cardona:2022mdq,PhysRevLett.131.111001}.

While a few groups argue that unaccounted for systematic errors could be responsible for observed phenomena or discrepancies in cosmological parameters (see, for instance, \cite{Efstathiou:2013via,Mazo:2022auo,Dainotti:2021pqg,Dainotti:2022bzg}), the possibility of new underlying physical theories has attracted more attention over the past years. The problem of explaining the current accelerating expansion of the Universe as well as the observed structures in the Cosmic Microwave Background (CMB) radiation and the matter distribution in the Large Scale Structure (LSS) is usually addressed by invoking new dynamical matter fields~\cite{Copeland:2006wr} and searching plausible modifications of GR~\cite{Clifton:2011jh}. A few examples of alternative models matching the $\Lambda$CDM background expansion while slightly differing in the evolution of linear order perturbations can be found in $f(R)$, Horndeski, and Scalar-Vector-Tensor (SVT) theories \cite{Arjona:2018jhh,Arjona:2019rfn,Cardona:2022lcz}.     
 
While scalar fields are usually preferred as a DE source when building cosmological models, vector fields are also found in the known matter and hence could indeed be plausible candidates for DE (see, for instance, Refs.~\cite{Geng:2021jso,Nakamura:2018oyy,deFelice:2017paw,DeFelice:2016yws,Armendariz-Picon:2004say,Koivisto:2008xf,Gomez:2020sfz,Alvarez:2019ues,Zuntz:2010jp,BeltranJimenez:2008iye,Bamba:2008ja,Bamba:2008xa}). A successful example in Ref.~\cite{Heisenberg:2020xak} invokes a Proca field and fits reasonable well available data. While the vector DE model in Ref.~\cite{Heisenberg:2020xak} slightly alleviates the tension in $H_0$, it does not help with the discrepancy in $\sigma_8$.  

Given the unknown nature of both dark matter and DE, it is worth to investigate a possible exchange of energy and momentum between them. A number of DE models featuring an interacting scalar field have been indeed proposed \cite{Amendola:1999er,Wang:2016lxa} which can provide plausible explanations. Vector DE models including an interaction with CDM are also found in the literature, but they are less common. In Ref.~\cite{Nakamura:2019phn} a coupled vector DE model was proposed so that the DE equation can reach values closer to a cosmological constant, thus alleviating tensions in similar preceding uncoupled vector DE models. The model in Ref.~\cite{Nakamura:2019phn} turns out to be a generalisation of the uncoupled Proca model in Ref.~\cite{Heisenberg:2020xak} which can provide an explanation for the current accelerating phase while partially alleviating the Hubble tension, even though it exacerbates the discrepancy in the strength of matter clustering~\cite{Heisenberg:2020xak}.   

In this work we investigate a vector DE model featuring an interaction with CDM. We disregard a coupling to baryonic matter as it seems to be disfavoured by observations ~\cite{Damour:1994zq}. The Lagrangian for the coupled vector DE model comes from Ref.~\cite{Nakamura:2019phn}  and we intend to study the model in light of current astrophysical measurements. Firstly, we provide the equations governing the background and the linear order perturbations which were missing in the literature (see Section \ref{Sec: Theoretical framework and phenomenology} and Appendices \ref{appendix:A}-\ref{appendix:C}). Secondly, by carrying out plausible approximations we manage to find analytical approximate solutions for the matter perturbations which turn out to be pretty similar to $\Lambda$CDM while having a slight dependence on the parameters of the model that may impact, for instance, the strength of matter clustering $\sigma_8$ (see Section \ref{Sec: Theoretical framework and phenomenology}). Thirdly, we compute cosmological constraints for the coupled vector DE model using available data sets. We assess whether or not the model can simultaneously alleviate tensions in the Hubble expansion rate and the strength of matter clustering which was thus far neglected in this context (see Sections \ref{section:cosmological-constraints}-\ref{section:conclusions}).       

\section{Theoretical framework and phenomenology}
\label{Sec: Theoretical framework and phenomenology}

The action for a generalised Proca theory also considering a coupling between the vector field and CDM reads\footnote{We define $\kappa \equiv 8 \pi G_N$, with $G_N$ being the bare Newton's constant. We use units where the speed of light $c=1$ as well as the short-hand notation $h_{\xi} \equiv \dfrac{dh}{d\xi}$ for the derivative of any free function $h$ with respect to a scalar $\xi$.}
\be
S = \int \text{d}^4 x \sqrt{-g} \left\{\frac{1}{2\kappa} R + G_{2}(X,F) + G_{3}(X) \nabla_{\mu} A^{\mu} + \[1 + Q f(X)\] \Lc  +\mc{L}_{\rm{SM}}\right\}.
\label{Eq: action}
\ee
In Eq.~\eqref{Eq: action} $g$ denotes the determinant of the metric $g_{\mu\nu}$ and $R$ the Ricci scalar; $\mc{L}_{\rm{SM}}$ is the Lagrangian of the Standard Model particles; the generalised Proca action for the vector field $A_\mu$ is defined through the free functions $G_2$ and $G_3$ which depend on the quadratic terms 
\begin{equation}
X \equiv -\frac{1}{2}A_{\mu}A^{\mu}, \qquad F \equiv -\frac{1}{4}F_{\mu\nu}F^{\mu\nu},
\label{Eq: definitions}
\end{equation}
with $F_{\mu\nu} \equiv \nabla_{\mu} A_\nu - \nabla_{\nu} A_\mu$ the strength tensor of the vector field; the Lagrangian of CDM $\Lc$ is coupled to the vector field through the function $f(X)$ considering a coupling strength given by the parameter $Q$. 

By varying  the action ~\eqref{Eq: action} with respect to $g^{\mu\nu}$, we obtain the gravitational field equations 
\be
G_{\mu\nu} = \kappa T_{\mu\nu},
\label{Eq: field equations}
\ee
where $G_{\mu\nu}$ is the Einstein tensor and $T_{\mu\nu}$ the energy-momentum tensor   
\begin{align}
T_{\mu\nu} &\equiv \(1 + Q f\) T^{(\rm{cdm})}_{\mu\nu} + Q f_X A_{\mu} A_{\nu} \Lc + G_{2 F} F_{\ \mu}^{\alpha} F_{\nu\alpha}  + 
G_{2} g_{\mu\nu} + G_{2 X} A_{\mu} A_{\nu} \nn \\ 
&+ G_{3 X}\[A_{\mu} A_{\nu} \nabla_{\alpha} A^{\alpha} + g_{\mu\nu} A^{\alpha} A^{\beta} \nabla_{\beta} A_{\alpha} - A^{\alpha} \(A_{\nu} \nabla_{\mu} A_{\alpha} + A_{\mu} \nabla_{\nu} A_{\alpha} \)\] + T_{\mu\nu}^{\rm{(SM)}}, \label{Eq: energy tensor}
\end{align}
with   
\be
T^{(\rm{cdm})}_{\mu\nu} \equiv - \frac{2}{\sqrt{-g}} \frac{\delta \(\sqrt{-g} \Lc \)}{\delta g_{\mu\nu}}, ~~~~ T^{(\rm{SM})}_{\mu\nu} \equiv - \frac{2}{\sqrt{-g}} \frac{\delta \(\sqrt{-g} \mc{L}_{\rm{SM}} \)}{\delta g_{\mu\nu}}, 
\label{Eq: energy tensor cdm}
\ee
respectively the CDM and Standard Model particles energy-momentum tensor.

The vector field equation of motion is computed by varying the action \eqref{Eq: action} with respect to the vector field $A_{\mu}$; we find
\ba
0 &= - Q f_{X} \Lc A^{\mu} + G_{2 F} \nabla_\nu F^{\nu\mu} + G_{2 X F} A^\nu \nabla_\alpha A_\nu F^{\mu \alpha} + G_{2 F F} \nabla^\mu A^\nu \nabla^\beta A^\alpha \nabla_\nu F_{\beta\alpha} \nn  \\
 &+ G_{2 F F} \nabla^\nu A^\mu \nabla^\beta A^\alpha \nabla_\nu F_{\alpha\beta} - G_{2 X} A^{\mu} - G_{3 X}\( A^{\mu} \nabla_{\alpha} A^{\alpha} - A^{\alpha} \nabla^{\mu} A_{\alpha} \). 
\label{Eq: vector eom}  
\ea
From Eqs. \eqref{Eq: energy tensor} and \eqref{Eq: vector eom}  energy-momentum conservation leads to 
\be
\nabla_{\mu}\[(1 + Q f)T^{\mu (\rm{cdm})}_{\ \nu} \] + Q f_{X} \Lc A^{\mu} \nabla_{\nu} A_{\mu} = 0,
\label{Eq: conservation eq}
\ee
namely, the effect of the interaction becomes apparent.

Since several observations support statistical homogeneity and isotropy on cosmological scales $\gtrsim 100\, \rm{Mpc}$  \cite{Planck:2018vyg, PhysRevLett.122.171301, Hogg:2004vw, Ade:2015hxq, Marinoni:2012ba, PhysRevLett.117.131302}, we assume that the geometry of the Universe is described by the Friedman-Lemaitre-Robertson-Walker (FLRW) metric in the Newtonian gauge
\begin{equation}
\text{d}s^{2} = a(\tau)^{2} \[ -\{ 1 + 2\Psi(\boldsymbol{x}, \tau) \} \text{d}\tau^{2} + \{1 -2\Phi(\boldsymbol{x}, \tau) \} \delta_{i j} \text{d}x^{i} \text{d}x^{j} \],
\label{Eq: FLRW}
\end{equation}
thus also allowing for tiny deviations from homogeneity which can be treated linearly. In Eq.~\eqref{Eq: FLRW} $a$ is the scale factor, $\tau$ the conformal time, and $\Psi$ and $\Phi$ are the gravitational potentials which depend on $\tau$ and the spatial coordinates $\boldsymbol{x}$. 

We also require the vector field to respect homogeneity and isotropy. Hence, we regard $A^\mu$ as having the following configuration 
\be
A^\mu = \( A^0(\tau) + \delta A^0(\boldsymbol{x}, \tau), \delta A^i(\boldsymbol{x}, \tau)  \),
\label{Eq: field profile}
\ee
with
\be
A^0 \equiv \frac{\phi(\tau)}{a}, \qquad \delta A^0 \equiv \frac{\delta\phi(\boldsymbol{x}, \tau)}{a}, \qquad \delta A^i \equiv  \delta^{i j} \frac{\partial_{j} \chi(\boldsymbol{x}, \tau)}{a^2},
\label{Eq: vector profile}
\ee
where $\phi$, $\delta \phi$, and $\chi$ are scalar quantities. Note the vector field only allows linear perturbations on the spatial part, neglecting any background contribution which could affect the isotropy assumption.  

In this work we investigate the coupled vector DE model defined by the functions \cite{Nakamura:2018oyy, Heisenberg:2020xak}
\begin{equation}
G_{2}(X, F) \equiv b_{2}X^{p_{2}} + F, \qquad G_{3}(X) \equiv b_{3} X^{p_{3}}, \qquad f(X) \equiv \(\frac{X}{X_{0}}\)^{q},
\label{Eq: model}
\end{equation}
where $b_2$, $p_2$, $b_3$, $p_3$, $q$, and $X_{0}$ are constants. Notice that in this case $G_{2 F} = 1, \, G_{2 F F} = 0, \, G_{2 X F} = 0$, hence Eqs.~\eqref{Eq: energy tensor} and \eqref{Eq: vector eom} get simplified.

\subsection{Background evolution}
\label{SubSec: Background evolution}

Disregarding linear perturbations, the vector field equation of motion \eqref{Eq: vector eom} becomes a constraint and reads
\begin{equation}
X\left(G_{2 X} + 3 \sqrt{2 X} \frac{\mathcal{H}}{a} G_{3 X} + \frac{Q f_{X}}{1 + Q f} \tildeLc \right) = 0,
\label{Eq: Back vector eom}
\end{equation}
where $\mathcal{H} \equiv a'/ a$ is the conformal Hubble parameter\footnote{A prime $'$ denotes derivative with respect to the conformal time $\tau$. Hubble parameter $H$ and conformal Hubble parameter are related through $\mathcal{H}=a H$.} and 
$\tildeLc \equiv ( 1 + Q f)\Lc$. Since we are not interested in the trivial solution $X = 0$, we focus on the solution given by the expression in parenthesis, namely, a constraint relating the functions $G_2$, $G_3$, and $f$. 

Inserting the constraint in Eq.~\eqref{Eq: Back vector eom} into the gravitational field equations \eqref{Eq: field equations} we obtain
\be 
\mathcal{H}^{2} = \frac{\kappa}{3} a^{2}\( \tilde{\rho}_{\rm{cdm}} + \rho_{\rm{DE}} + \rho_{\rm{SM}}\), \qquad 2 \mathcal{H}' + \mathcal{H}^{2} = - \kappa a^{2} \(P_{\rm{DE}}+P_{\rm{SM}}\), 
\label{Eq: Friedman eqs} 
\ee
where we consider
\be
\trhoc = - \tildeLc = (1 + Q f )\rho_\mathrm{cdm}, \qquad \rho_{\rm{DE}} = -G_{2}, \qquad P_{\rm{DE}} = G_{2} - \frac{\phi^{2}\phi'}{a} G_{3 X},
\label{Eq: densities}
\ee
$\rho_{\rm{cdm}}$ being the density of the bare CDM; $\rho_{\rm{DE}}$ and $P_{\rm{DE}}$ denote the DE density and pressure, respectively, from which we can compute the DE equation of state $w_{\rm{DE}} \equiv P_{\rm{DE}}/\rho_{\rm{DE}}$;  $\rho_{\rm{SM}}$ and $P_{\rm{SM}}$ denote contributions to density and pressure from the Standard Model particles (i.e., baryons, photons, neutrinos). Here  $\trhoc$ is an effective CDM density and the choices $\Lc = -\rho_{\rm{cdm}}$, $\tildeLc = - \trhoc$, are consistent with the form of the Lagrangian of a pressure-less perfect fluid (see, for instance, Refs. \cite{Koivisto:2005nr, Avelino:2018qgt}).

Note that the coupling terms in the definition of the energy-momentum tensor ~\eqref{Eq: energy tensor} hinder a clear identification of both CDM and DE contributions. We can overcome this difficulty after obtaining the Friedmann equations (see Appendix \ref{appendix:A} for details). By computing the time derivatives of $\trhoc$ and $\rhoDE$ in Eq.~\eqref{Eq: densities}  
\be
\tilde{\rho}'_{\rm{cdm}} + 3 \mathcal{H} \trhoc = \frac{\phi \phi' Q f_{X}}{1 + Q f} \tilde{\rho}_{\rm{cdm}}, \qquad \rho_{\mathrm{DE}}' + 3\mathcal{H}\( \rho_{\mathrm{DE}} + P_{\mathrm{DE}} \) = -\frac{\phi \phi' Q f_{X}}{1 + Q f} \tilde{\rho}_{\rm{cdm}},
\label{Eq: continuity eqs}
\ee
hence it becomes clear the effective CDM and DE are exchanging energy through the interaction term in the right-hand side. In Eqs.~\eqref{Eq: continuity eqs} we assumed CDM $\rho_\mathrm{cdm}$ obeys the usual continuity equation $\rho'_\mathrm{\rm{cdm}} = - 3 \mc{H} \rho_\mathrm{\rm{cdm}}$. 

\begin{figure*}[h]
\centering
\includegraphics[width=\textwidth]{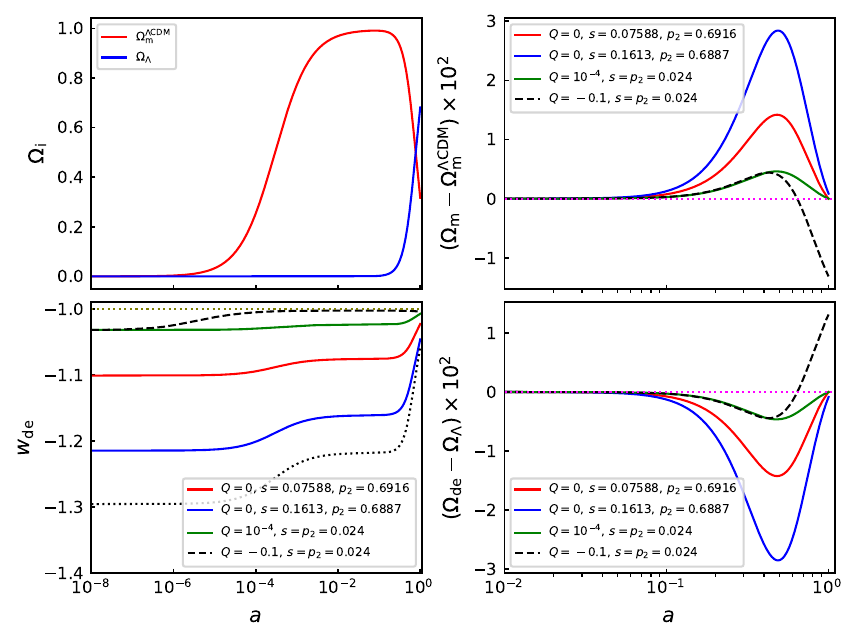}
\caption{\textit{Upper, left panel:} Evolution of the density parameters in the standard cosmological model $\Lambda$CDM. \textit{Upper, right panel:} Difference between effective matter density parameter in coupled vector DE model and $\Lambda$CDM. \textit{Lower, left panel:} DE equation of state as compared to a cosmological constant (olive, dotted line) and the best fit of our baseline result depicted as a black, dotted curve (corresponding parameters given in caption of Fig.~\ref{Fig:fs8}). \textit{Lower, right panel:} Difference between DE density parameter in coupled vector DE model and $\Lambda$CDM. While the uncoupled model generally has a phantom DE equation of state, a non vanishing negative coupling in the dark sector can feature a similar cosmological constant behaviour in the DE equation of state at late times. Other cosmological  parameters are fixed to the $\Lambda$CDM baseline result reported by the Planck Collaboration~\cite{Planck:2018vyg}.}
\label{Fig:Background}
\end{figure*}

As shown in Fig.~\ref{Fig:Background}, the coupling between CDM and DE indeed implies some relevant modifications to the background universe, which become apparent in the evolution of density parameters and DE equation of state. These changes can be understood by taking into account that Eq.~\eqref{Eq: Back vector eom} allows us to write $G_{3 X}$ as
\be
G_{3 X} = -\frac{a}{3 \mc{H} \phi}(1 - r_{\rm Q}) G_{2 X}, \qquad r_{\rm{Q}} \equiv \frac{Q f_X}{G_{2 X}} \rho_{\rm{cdm}},
\label{Eq: G3X in Q}
\ee
so that the DE pressure in Eq.~\eqref{Eq: densities} can differ from a cosmological constant depending on the field evolution. Eqs.~\eqref{Eq: G3X in Q} make apparent the relation between the free functions $G_2$ and $G_3$ and the coupling function $f$.
  
The variable $r_{\rm{Q}}$ in Eq.~\eqref{Eq: G3X in Q} allows us to find an equation governing the evolution of the vector field. Firstly, we derive the vector field equation \eqref{Eq: Back vector eom} with respect to $\tau$ and use Eqs.~\eqref{Eq: Friedman eqs} and \eqref{Eq: G3X in Q} to eliminate $\mathcal{H}'$ and $G_{3 X}$, respectively. Secondly, we take advantage of the polynomial form in the functions \eqref{Eq: model} to simplify $G_{2 X X}$ and $G_{3 X X}$. Finally, we use Eq.~\eqref{Eq: continuity eqs} to obtain
\be
\epsilon_{\phi} \equiv \frac{\rho_{\rm{DE}}'}{\mathcal{H} \rho_{\rm{DE}}} = \frac{3(1 + r_{\rm{Q}}) + (1 - r_{\rm{Q}})(\kappa a^{2}\mc{H}^{-2}P_{\rm{SM}}-3\Omega_{\rm{DE}})}{(1 + s_{1} r_{\rm{Q}}) + s(1 - r_{\rm{Q}})^{2} \Omega_{\rm{DE}}} s, 
\label{Eq: eps_phi}
\ee
where $\Omega_\text{DE} \equiv \kappa a^{2}\rho_\text{DE}/3\cH^{2}$ is the DE density parameter and we defined the parameters
\be
s = \frac{p_{2}}{p}, \qquad s_{1} = \frac{2q -2p_{3} - 1}{p}, \qquad p = 2p_{3} - 2p_{2} + 1.
\label{Eq:param-def}
\ee
Note that, as we show in Appendix \ref{appendix:C}, in order to keep the system stable the parameter $q$ is determined by $p_3$ and $p_2$ as $q=2p_{3}-p_{2}+1$ which in turn yields $s_1=1$.

\subsection{Linear order perturbations}
\label{subsection:perturbations}

In this work we focus on scalar perturbations. From Eq.~\eqref{Eq: vector eom} we compute the temporal component of the vector field equation
\ba  
& & \[a\phi^{2}G_{2XX} +3\cH\phi\(\phi^{2}G_{3XX}+G_{3X}\)-a\phi^{2}Qf_{XX}\rho_{\rm{cdm}}\]\(\delta\phi+\phi\Psi\) - k^{2}a^{-1}\phi G_{3X}\chi \nn \\
& & + k^{2}a^{-2}\[a\(\delta\phi+2\phi\Psi\)+\chi' \] -a\phi Qf_{X}\delta\rho_{\rm{cdm}}-3\phi^{2}G_{3X}\(\cH\Psi+\Phi'\) =0, 
\label{eq:perturbed_vector_field_equation_0}
\ea
as well as the spatial component 
\be 
k^{2}a^{-2}\{\partial_{\tau}\[a\(\delta\phi+2\phi\Psi\)\]+\chi'' \}  
+ k^{2}\phi G_{3X}\(\delta\phi+\phi\Psi\)+k^{2}a^{-1}\phi'G_{3X}\chi = 0.
\label{eq:spatial-perturbation-vector-field-equation}
\ee
The time-time component of the linearised Einstein field equations reads  
\ba
k^{2}\Phi &+& 3\mathcal{H}\(\mathcal{H}\Psi+\Phi'\) = -\frac{\kappa}{2}  a^{2}\left\{ \delta\rho_{\rm{SM}}  +\[1+Qf(X)\]\delta\rho_{\rm{cdm}} -\phi^{2}Qf_{X}\delta\rho_{\rm{cdm}} \right.  \nn \\
&+&\[\phi^{2}G_{2XX} +3a^{-1}\cH\phi\(\phi^{2}G_{3XX}+2G_{3X}\)-\phi^{2}Qf_{XX}\rho_{\rm{cdm}}\]\(\phi\delta\phi+\phi^{2}\Psi\)  \nn \\
&-& \left.  k^{2}a^{-2}\phi^{2}G_{3X}\chi -3 a^{-1}\phi^{3}G_{3X}\(\cH\Psi+\Phi' \) \right\},
\label{eq:perturbed_Friedmann_0}
\ea 
the time-space components 
\ba
k^{2}\(\cH\Psi +\Phi' \) &=& \frac{\kappa}{2} a^{2}\left\{\(\rho_{\rm{SM}} + P_{\rm{SM}}\)\theta_{\rm{SM}} + \[1+Qf(X)\]\rho_{\rm{cdm}}\theta_{\rm{cdm}}   \right. \nn \\
&-& \left. k^{2}a^{-1}\phi^{2}G_{3X}\(\delta\phi+\phi\Psi\)\right\},
\label{eq:perturbed_Friedmann_0i}
\ea
the space-space components
\ba
&\Phi''& + \cH\( \Psi' + 2\Phi'\)+\(\cH^{2}+2\cH'\)\Psi + \frac{1}{3}k^{2}\(\Phi-\Psi\) =  \frac{\kappa}{2} a^{2} \{\delta P_{\mathrm{SM}}  \nn \\\
&+&\phi G_{2X}\(\delta\phi+\phi\Psi\)-a^{-1}\partial_{\tau}\[\phi^{2}G_{3X}\(\delta\phi+\phi\Psi\)\] + a^{-1}\phi^{2}\phi'G_{3X}\Psi\},
\label{eq:perturbed_Friedmann_ii}
\ea
and the off-diagonal space-space components
\be
k^{2}\(\Phi-\Psi\) = \frac{3}{2}\kappa a^{2} \sum_i \(\rho_i+P_i\)\sigma_i,
\label{eq:perturbed_Friedmann_ik}
\ee
where $i$ stands for Standard Model components and $\sigma_i$ refers to their anisotropic stress. We consider CDM as having a vanishing anisotropic stress $\sigma_{\rm{cdm}}=0$. Note that from our definition of the energy momentum tensor Eq.~\eqref{Eq: energy tensor} and our choice for the field configuration Eq.~\eqref{Eq: field profile}, the DE-CDM coupling does not induce any additional anisotropic stress. In Eqs.~\eqref{eq:perturbed_Friedmann_0}-\eqref{eq:perturbed_Friedmann_ik} $\delta\rho_{\mathrm{SM}}, \theta_{\rm{SM}}$ and $\delta P_{\rm{SM}}$ are respectively the density perturbations, the velocity divergence, and the pressure perturbation of the Standard Model particles; $\delta\rho_{\rm{cdm}} = -\delta T^{0(\rm{cdm})}_{0}$; $ik_{i}\delta T^{0(\rm{cdm})}_{~i} = \rho_{\rm{cdm}}\theta_{\rm{cdm}}$.    

Note that from Eqs.~\eqref{eq:perturbed_Friedmann_0}-\eqref{eq:perturbed_Friedmann_ii} we can define DE perturbations 
\be 
\(\rho_{\mathrm{DE}}+P_{\mathrm{DE}}\)\theta_{\mathrm{DE}} \equiv  -k^{2}a^{-1}\phi^{2}G_{3X}\(\delta\phi+\phi\Psi\),
\label{eq:DE-velocity-divergence}
\ee 
\ba
\delta\rho_{\mathrm{DE}} &\equiv& \[\phi^{2}G_{2XX} +3a^{-1}\mathcal{H}\phi\(\phi^{2}G_{3XX}+2G_{3X}\)-\phi^{2}Qf_{XX}\rho_{\rm{cdm}}\]\(\phi\delta\phi+\phi^{2}\Psi\)\nn \\
&-& k^{2}a^{-2}\phi^{2}G_{3X}\chi -3 a^{-1}\phi^{3}G_{3X}\(\mathcal{H}\Psi+\Phi' \) -\phi^{2}Qf_{X}\delta\rho_{\mathrm{cdm}},
\label{eq:DE-density-perturbation}
\ea
\be
\delta P_{\rm{DE}} \equiv \phi G_{2X}\(\delta\phi+\phi\Psi\) -a^{-1}\partial_{\tau}\left[\phi^{2}G_{3X}\(\delta\phi+\phi\Psi\)\right]+a^{-1}\phi^{2}\phi'G_{3X}\Psi,
\ee
and effective CDM perturbations 
\be
\tilde{\rho}_{\rm{cdm}}\tilde{\theta}_{\rm{cdm}} \equiv \left[1+Qf(X)\right]\rho_{\rm{cdm}}\theta_{\rm{cdm}},
\label{eq:DM-velocity-divergence}
\ee
\be
\delta\tilde{\rho}_{\rm{cdm}} \equiv \left[1+Qf(X)\right]\delta\rho_{\rm{cdm}}.
\label{eq:DM-density-perturbation}
\ee 
Following Ref.~\cite{Heisenberg:2020xak} we define the auxiliary variables
\begin{equation}
\delta_{\chi} \equiv \frac{k\chi}{a\phi},~~~~~\delta_{\phi} \equiv \frac{k(\delta\phi+\Psi\phi)}{\dot{\phi}},
\label{eq:delta_phi}
\end{equation}
\begin{equation}
\mathcal{Z} \equiv -\frac{k^{2}\phi}{a^{3}\rho_{\mathrm{DE}}}\left(a(\delta\phi+2\Psi\phi) +\partial_{\tau}\chi\right),
\label{eq:z}
\end{equation}
\be
\mathcal{Q} \equiv -\(\frac{\mathcal{Z}}{p}+\frac{2}{3}\frac{s k}{\mathcal{H}}(1-r_{\rm Q})\delta_{\chi}\),
\label{eq:auxiliar_3}
\ee
which help to simplify the system of differential equations. Using Eqs.~\eqref{eq:delta_phi}-\eqref{eq:z} the perturbed vector field equations of motion \eqref{eq:perturbed_vector_field_equation_0}-\eqref{eq:spatial-perturbation-vector-field-equation} become
\be
-\mathcal{Z} + p\(1+s_{1}r_{\rm Q}\) \frac{\cH}{k}\epsilon_{\phi}\delta_{\phi}-\frac{2}{3}\frac{k}{\cH}p_{2}(1-r_{\rm{Q}})\delta_{\chi} + 2p_{2}r_{\rm Q}\delta_{\rm{cdm}}-\frac{2p_{2}}{\cH}(1-r_{\rm Q})(\cH\Psi+\Phi') = 0,
\label{eq:auxiliar_1}
\ee
\be
-\mathcal{Z}' +\mathcal{H}\mathcal{Z}\[\epsilon_{\phi}\(\frac{1}{2p_{2}}-1\)-3 \]+\frac{1}{2}\frac{k}{\mathcal{H}}(1-r_{\rm Q})\epsilon_{\phi}\(\delta_{\phi}+\delta_{\chi}\) = 0,
\label{eq:auxiliar_2}
\ee
where we used that $\tilde{\delta}_{\rm{cdm}}=\delta_{\rm{cdm}}$ defined as $\tilde{\delta}_{\rm{cdm}} \equiv \delta\tilde{\rho}_{\rm{cdm}}/\tilde{\rho}_{\rm{cdm}}$. The derivative of $\delta_{\chi}$ is likewise given by 
\be
\delta_{\chi}' = -\(\frac{\epsilon_{\phi}}{2p_{2}}+1\)\mathcal{H}\delta_{\chi}-\frac{\epsilon_{\phi}}{2p_{2}}\mathcal{H}\delta_{\phi}-\frac{a^{2}\rho_{\mathrm{DE}}}{k\phi^{2}}\mathcal{Z}-k\Psi.
\label{eq:auxiliar_3}
\ee
The vector field equations of motion ~\eqref{eq:auxiliar_1}-\eqref{eq:auxiliar_2} can then be rewritten as
\be
\mathcal{Q}'  = -2\(1-\frac{3}{4}\alpha\)\mathcal{H}\mathcal{Q}+\left[s\mathcal{R}_{1}+3\(1+s\mathcal{R}_{1}\)c_{A}^{2}\right]\mathcal{H}\mathcal{Z} + \frac{2}{3}\frac{k^{2}s}{\mathcal{H}}(1-r_{\rm Q})\Psi,
\label{Eq:Q-prime}
\ee
\ba
\mathcal{Z}' &=& 3\frac{r_{\rm Q}+w_{\mathrm{DE}}}{1-r_{\rm Q}}\mathcal{H}\mathcal{Z}-\(\frac{k^{2}}{3\mathcal{H}^{2}}\mc{R}_{3} + \frac{3}{2}\alpha\)\mathcal{H}\mathcal{Q}+ \frac{2}{3}\frac{k^{2}s}{\mathcal{H}^{2}}(1-r_{\rm{Q}})\mc{R}_{3}(\mathcal{H}\Psi+\Phi') \nn \\
&-& \frac{2}{3}\frac{k^{2}s}{\mathcal{H}}\mc{R}_{3}r_{\rm Q}\delta_{\mathrm{cdm}},
\label{Eq:Z-prime}
\ea
where
\be
c_{A}^{2}\equiv p^{-1}\(1+s\mathcal{R}_{1}\)^{-1}\left[\frac{2sp}{3\phi^{2}_{pl}}\mathcal{R}_{1} + \frac{1}{3}\(1-sp\mathcal{R}_{1}\)+\frac{1}{2}\(1+2s-\frac{1}{p}\)\alpha \right],
\ee
\be
\mathcal{R}_{1} \equiv (1-r_{\rm{Q}})\Omega_{\rm{DE}}, \qquad \phi_{pl} \equiv \frac{\phi}{M_{\mathrm{pl}}}, \qquad \alpha \equiv \frac{\epsilon_{\phi}}{3s}.
\ee
DE perturbations ~\eqref{eq:DE-velocity-divergence}-\eqref{eq:DE-density-perturbation} can be likewise rewritten as 
\begin{equation}
\(\rho_{\rm{DE}}+P_{\rm{DE}}\)\theta_{\rm{DE}} = \(1+s\mathcal{R}_{2}\)^{-1}\left\{\frac{k^{2}}{3\mc H}\rho_{\rm{DE}}\mc Q - s\mc R_{2}\(\rho+P\)\theta  + \frac{2}{3}sk^{2} \frac{r_{\rm Q}}{\mc H}\mc R_{3}\rho_{\rm{DE}}\delta_{\rm{cdm}} \right\},
\label{eq:DE-velocity-perturbation}
\end{equation}
\begin{equation}
\delta\rho_{\rm{DE}} = \rho_{\rm{DE}}\mc Z -\frac{3\mc H}{k^{2}}\(\rho_{\rm{DE}} + P_{\rm{DE}}\)\theta_{\rm{DE}},
\end{equation}
with the parameters $\mc R_{2}$ and $\mc R_{3}$  
\be
\mc R_{2} \equiv \mc R_{3} \mc R_{1}, \qquad \mc R_{3} \equiv \frac{1-r_{\rm{Q}}}{1+s_{1}r_{\rm{Q}}}.
\ee
In Eq.~\eqref{eq:DE-velocity-perturbation} the term $(\rho + P)\theta$ refers to the sum of Standard Model particles contributions and effective CDM. 

Concerning the equations governing the evolution of effective CDM perturbations, we obtain them by taking the spatial ($\nu = i$) and temporal ($\nu = 0$) components in Eq.~\eqref{Eq: conservation eq}, namely, 
\be
\theta'_{\rm{cdm}}  +\mathcal{H}\theta_{\rm{cdm}} - k^{2}\Psi = \zeta\mc H \epsilon_{\phi}\(k \delta_{\phi}-\theta_{\rm{cdm}}\)
\label{eq:Euler_for_cdm},
\ee
\be
\delta'_{\rm{cdm}} + \theta_{\rm{cdm}} -3\Phi' = 0.
\label{eq:perturbed_energy_for_cdm}
\ee
In the appendix \eqref{appendix: Euler_equations}, we show that DE and effective CDM as we define them satisfy the conservation of energy and momentum. A second order differential equation for $\delta_{\rm{cdm}}$ is obtained by taking the derivative of Eq.~\eqref{eq:perturbed_energy_for_cdm} and replacing $\theta_{\rm{cdm}}$ from Eq.~\eqref{eq:Euler_for_cdm} 
\be
\delta''_{\rm{cdm}} + \mc H\(1 + \zeta\epsilon_{\phi}\)\delta'_{\rm{cdm}} = 3\(\mc H \Phi' +\Phi''\)-k^{2}\Psi+\mc H\zeta\(3\Phi'-k\delta_{\phi}\).
\ee
The parameter $\zeta\epsilon_{\phi} \equiv \frac{q}{p_{2}}\frac{Qf}{1 +Qf}\epsilon_{\phi}$ is a new drag term due to the energy exchange in the dark sector~\cite{Asghari:2019qld}. The quantity $\zeta\mc H$ makes apparent the competition between the interaction and the Universe's expansion rate. 
\begin{figure*}[h]
\centering
\includegraphics[width=\textwidth]{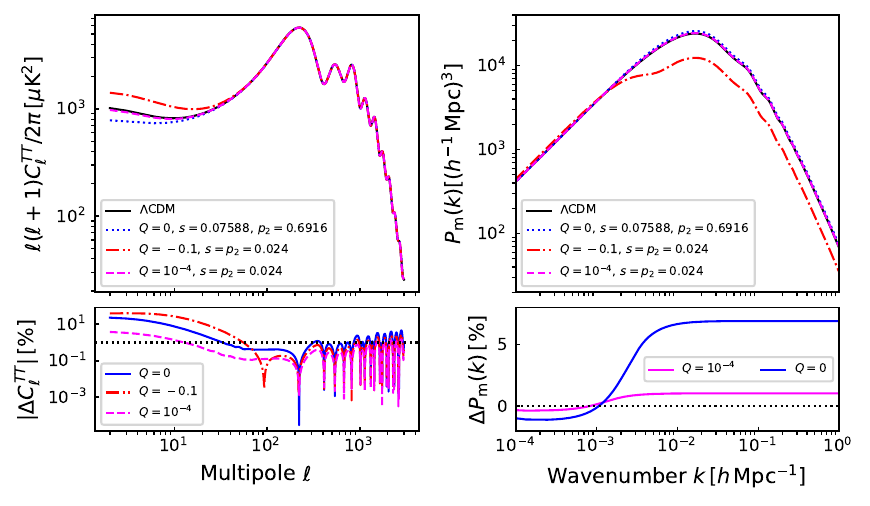}
\caption{\textit{Upper panels:} CMB TT angular power spectrum (left) and matter power spectrum at $z=0$ (right). We show the result for the standard cosmological model along with a few examples of uncoupled and coupled vector DE model (see also green and black dashed curves in Fig.~\ref{Fig:Background} for background behaviour). As in Fig.~\ref{Fig:Background}, parameters not shown in the legend are fixed to the baseline $\Lambda$CDM result reported by the Planck Collaboration~\cite{Planck:2018vyg}. \textit{Lower panels:} We display the percentage difference of the quantities plotted in the upper panels relative to the standard cosmological model.}
\label{Fig:linear-perturbations}
\end{figure*}
We implemented the model \eqref{Eq: model} in the popular Boltzmann solver \texttt{CLASS} ~\cite{Blas:2011rf} and managed to solve the full system of differential equations governing both background and linear order perturbations. Figure~\ref{Fig:linear-perturbations}  shows the solutions [i.e., temperature CMB angular power spectrum $C_\ell^{\rm TT}$ and matter power spectrum $P(k,z=0)$] for a few representative models including $\Lambda$CDM, an uncoupled vector field ($Q=0$), and a coupled vector DE model. We can therefore test the predictions of the model against available data sets. However, before proceeding with the statistical analysis we further study the solutions on the matter dominance regime so that we can have an insight into the predictions of the coupled vector DE model.      

\subsubsection{CDM perturbations in the matter dominance regime}

\begin{figure*}[h]
\centering
\includegraphics[width=\textwidth]{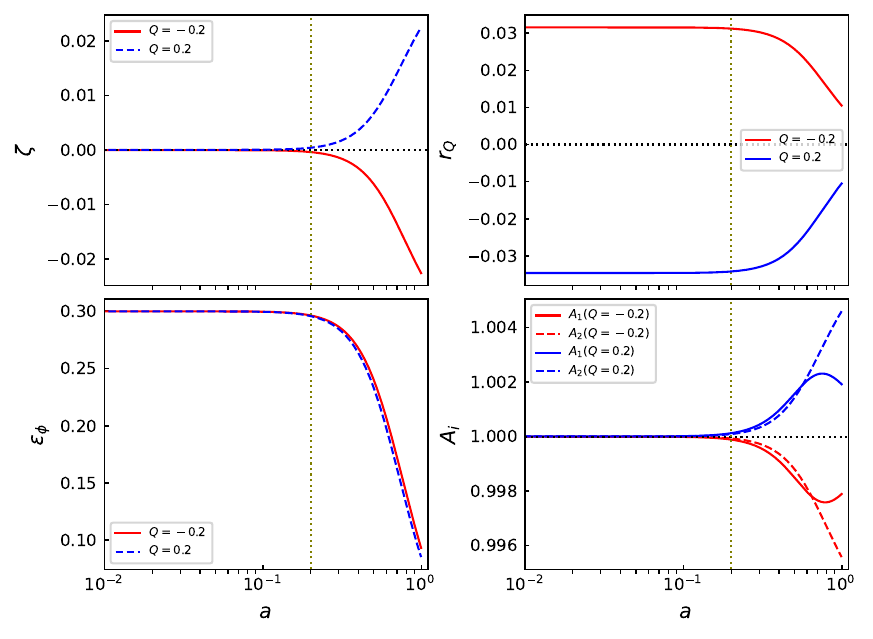}
\caption{The behaviour of a few quantities affecting the evolution of linear order perturbations in the coupled vector DE model. We use $s=0.1$ and $p_2=0.6$. Besides the parameters showed in the panels, we fix other relevant cosmological parameters to the Planck Collaboration baseline result~\cite{Planck:2018vyg}.}
\label{A1_and_A2}
\end{figure*}

As we can see in Fig.~\ref{Fig:Background}, the background evolution of our coupled vector DE model can be quite similar to $\Lambda$CDM. Therefore, it is reasonable to simplify the equations governing the background evolution during matter dominance, namely, 
\begin{equation}
\cH^{2}\approx H_{0}^{2}\Omega_{\mathrm{cdm}}^{0} a^{-1},
\label{Eq:HMD}
\end{equation}
where $\Omega_{\mathrm{cdm}}^{0}$ is constant and $\Omega_{\mathrm{DE}}\approx 0$. Since we assume CDM has a vanishing anisotropic stress, we can further simplify our equations: from Eq.~\eqref{eq:perturbed_Friedmann_ik} the gravitational potentials satisfy $\Psi = \Phi$ after matter dominance onset. As a result of our assumptions,
\begin{equation}
\mathcal{R}_{1}\approx 0,~~~~\epsilon_{\phi}\approx 3s,~~~~w_{\rm DE}\approx-1-s(1-r_{\rm{Q}}),~~~~\alpha\approx 1,~~~~c_{\mathrm{A}}^{2}\approx \frac{1}{p}\left(\frac{5}{6}+s-\frac{1}{2p}  \right),
\label{Eq:MD}
\end{equation}
and we can write the system of differential equations for both DE [i.e., Eqs.~\eqref{Eq:Q-prime}-\eqref{Eq:Z-prime}] and CDM [i.e., Eqs.~\eqref{eq:Euler_for_cdm}-\eqref{eq:perturbed_energy_for_cdm}] perturbations during matter dominance as 
\be
\mathcal{Z}' - 3\mathcal{H}\frac{r_{\rm{Q}}+w_{\mathrm{DE}}}{1-r_{\rm{Q}}}\mathcal{Z} = -\left(\frac{k^{2}}{3\mathcal{H}^{2}}\mathcal{R}_{3} +\frac{3}{2}\right)\mathcal{H}\mathcal{Q} +\frac{2}{3}\frac{sk^{2}}{\mathcal{H}^{2}}\left(1-r_{\rm{Q}}\right)\mc{R}_{3}\mathcal{H}\Phi  
-\frac{2}{3}\frac{sk^{2}}{\mathcal{H}}\mathcal{R}_{3}r_{\rm{Q}}\delta_{\rm{cdm}}, 
\label{eq: z_prime_matter}
\ee
\begin{equation}
\mathcal{Q}' + \frac{1}{2}\mathcal{H}\mathcal{Q} = 3c_{\mathrm{A}}^{2}\mathcal{H}\mathcal{Z}    
 + \frac{2}{3}\frac{k^{2}}{\mathcal{H}}(1 -r_{\rm{Q}})s\Phi,
\label{eq: Q_prime_matter}
\end{equation}
\begin{equation}
\delta_{\mathrm{cdm}}' = -\theta_{\mathrm{cdm}},
\label{eq: delta_m_prime}
\end{equation}
\begin{equation}
\theta_{\mathrm{cdm}}' + \mathcal{H}\(1+\zeta\epsilon_{\phi}\)\theta_{\mathrm{cdm}} = k^{2}\(1+2s \zeta \mathcal{R}_{3}\)\Phi 
- \frac{k^{2}}{1-r_{\rm{Q}}}\zeta\(\mc{Q}+2 s r_{\rm{Q}}\delta_{\rm{cdm}}\).
\label{Eq: theta_m_prime}
\end{equation}
Note that in Eqs.~\eqref{eq: z_prime_matter}-\eqref{Eq: theta_m_prime} we took into account the quasi-static approximation by neglecting the temporal evolution of the gravitational potential. As we will see below, our numerical experiments indicate indeed that $\Phi' \approx 0$ is a quite good approximation during the matter dominance regime. Moreover, we also found that for reasonable values of parameters, $\zeta \approx 0$ during matter dominance (see Fig.~\ref{A1_and_A2}). Consequently, the last term in Eq.~\eqref{Eq: theta_m_prime} does not significantly contribute to the evolution and can be safely neglected in our search for analytical approximate solutions. We simplify Eq.~\eqref{Eq: theta_m_prime}   
\be
\frac{d\theta_{\mathrm{cdm}}}{dN}=-A_{1}\theta_{\mathrm{cdm}} +k^{2}A_{2}\Phi,
\label{eq: theta_m_prime}
\ee
where $N \equiv \ln a$ denotes e-folds number, $A_{1} \equiv 1 +\zeta\epsilon_{\phi}$, and $A_{2} \equiv 1+2s\zeta \mathcal{R}_{3}$. In Fig.~\ref{A1_and_A2} we depict $\zeta$, $r_{\rm{Q}}$, $\epsilon_\phi$, $A_1$, and $A_2$ for plausible parameter values. Since during matter dominance the vector field is subdominant, the coupling function $f(X)\ll 1$, and $\zeta \approx 0$. As a result, we expect $A_1$ and $A_2$ to be approximately constant.

Under the assumptions above, CDM perturbations are uncoupled to DE perturbations and we can actually find analytical approximate solutions $\delta_{\rm{cdm}}$ and $\theta_{\rm{cdm}}$. They read 
\be
\theta_{\rm{cdm}} = -\delta_0 H_0 \sqrt{\Omega_{\rm{cdm}}^0} \sqrt{a} + \Theta_{1} a^{-A_{1}},
\label{eq:CDM-solutions-theta}
\ee
\be
\delta_{\rm{cdm}} = \delta_0 a + \delta_1 - \Theta_1 \frac{2 a^{(\frac{1}{2}-A_{1})}}{H_0 \sqrt{\Omega_{\rm{cdm}}^0} (1 - 2 A_1)},
\label{eq:CDM-solutions-delta}
\ee
where  $\Phi_{0}$ is the constant value of the gravitational potential, $\Theta_{1}$ is a constant of integration and 
\be 
\delta_0 \equiv - \frac{2 A_2 k^2 \Phi_0}{H_0^2 \Omega_{\rm{cdm}}^0 (1 + 2 A_1)}, \qquad 
\delta_1 \equiv \frac{3 \delta_0 H_0^2 \Omega_{\rm{cdm}}^0}{k^2},
\ee
are constants. Note that CDM solutions~\eqref{eq:CDM-solutions-theta}-\eqref{eq:CDM-solutions-delta} might lead to growing modes absent in the $\Lambda$CDM solutions when $A_1<0$ and $A_1 \rightarrow 1/2$. However, since the vector field is subdominant during the matter dominated epoch, $A_1$ and $A_2$ seem to remain positive and  $A_1 \approx A_2 \approx1$ for reasonable values of parameters (see Fig.~\ref{A1_and_A2}). Therefore, as long as $A_1>1/2$ we can safely neglect the last terms in Eqs.~\eqref{eq:CDM-solutions-theta}-\eqref{eq:CDM-solutions-delta} by effectively setting $\Theta_1=0$. We obtain
\be
\theta_{\rm{cdm}} = -\delta_0 H_0 \sqrt{\Omega_{\rm{cdm}}^0} \sqrt{a}, 
\label{eq:CDM-solutions-theta-2}
\ee
\be
\delta_{\rm{cdm}} = \delta_0 a + \delta_1 
\label{eq:CDM-solutions-delta-2}.
\ee
Finally, note that in the special case $A_1=A_2 \rightarrow 1$, Eqs.~\eqref{eq:CDM-solutions-theta-2}-\eqref{eq:CDM-solutions-delta-2} fully match the well known solutions of matter perturbations in the standard cosmological model (see, for instance, \cite{Sapone:2009mb}).  

\begin{figure*}[h]
\centering
\includegraphics[width=\textwidth]{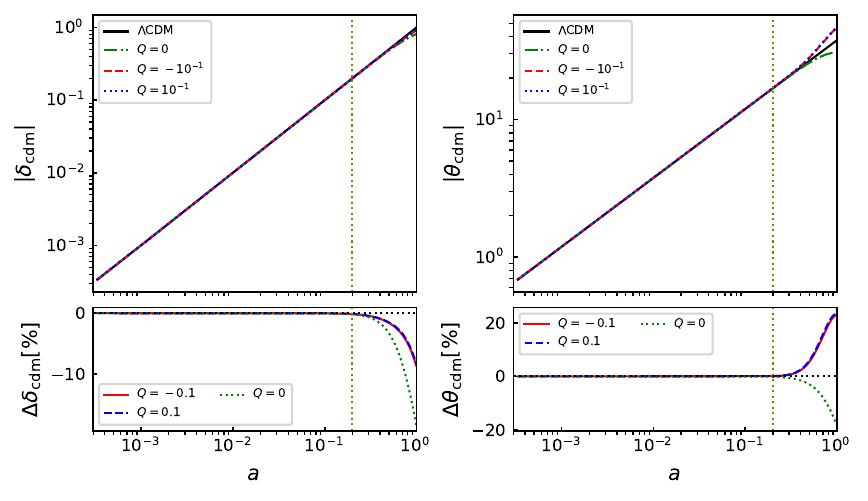}
\caption{\textit{Upper panels:} Numerical solutions for CDM perturbations in the coupled vector DE model. The black solid curve shows the analytical solution in the matter dominated epoch for the $\Lambda$CDM model. We depict the mode  $k=400H_{0}$ and solve the differential equations using $\delta_0=1$, $s=0.1$, and $p_2=0.6$. Other relevant parameters match the Planck Collaboration baseline result. \textit{Lower panels:} We show the percentage difference of our numerical solutions with respect to the analytical solutions for $\Lambda$CDM. Olive vertical dotted lines approximately indicate the end of CDM dominance in the standard cosmological model.}
\label{delta_theta}
\end{figure*}

In Fig.~\ref{delta_theta} we compare the approximate analytical solutions~\eqref{eq:CDM-solutions-theta-2}-\eqref{eq:CDM-solutions-delta-2} to the full numerical solution of the system of differential equations \eqref{Eq:Q-prime}-\eqref{Eq:Z-prime}, \eqref{eq:Euler_for_cdm}-\eqref{eq:perturbed_energy_for_cdm}. The agreement during matter dominance is quite good and solutions differ only when the assumptions break down, i.e., when DE starts to be the dominant component recently (see also upper panels in Fig.~\ref{Fig:perturbations-CLASS}).   

\subsubsection{DE perturbations during matter dominance under the sub-horizon approximation}

\begin{figure*}[h]
\centering
\includegraphics[width=\textwidth]{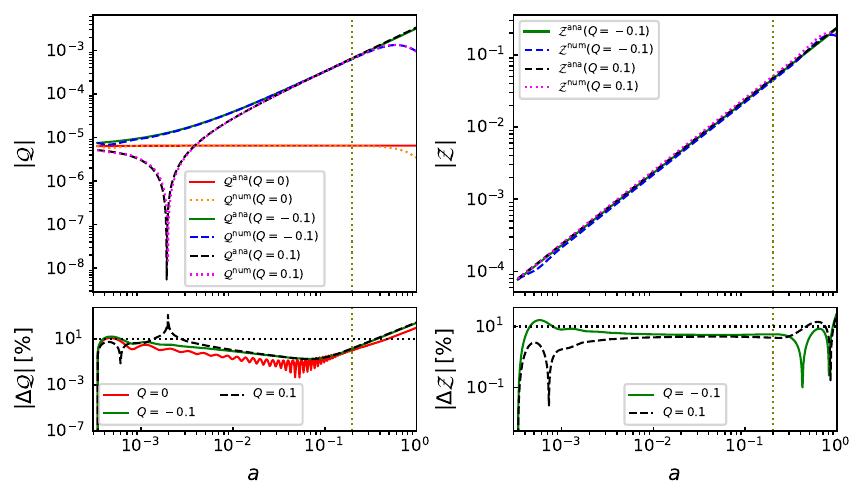}
\caption{\textit{Upper panels:} Analytical approximate solutions under the sub-horizon approximation for the variables $\mc{Q}$ and $\mc{Z}$ related to DE perturbations. During matter dominance analytical solutions closely follow the numerical solution, they deviate at the onset of the DE domination epoch. We display a mode $k=400H_{0}$ and solve the differential equations using $s=0.1$, $p_2=0.6$, and $\delta_0=1$. Other parameters are fixed to the Planck Collaboration baseline result. \textit{Lower panels:} We show the percentage difference of our analytical solutions with respect to the numerical solutions. Olive vertical dotted line approximately indicates the end of CDM dominance in the standard cosmological model.}
\label{Q_Z}
\end{figure*}

We could only find accurate analytical approximate solutions for DE perturbations in the matter dominated epoch after applying the sub-horizon approximation, namely, we regard scales satisfying  $k/\mathcal{H}\gg1$. First, let us consider Eq.~\eqref{eq: Q_prime_matter} for 
$\mathcal{Q}$. In order to avoid an uncontrolled growth of DE perturbations on small scales, we require the terms on the right hand side to cancel out. As a result we find 
\be
\mc{Z} =  \frac{2 k^2 s (r_{\rm{Q}} - 1) \Phi_{0}}{9 c_{A}^{2} \mathcal{H}^2}.
\label{eq:DE-solution-Z}
\ee
Taking into consideration that during matter dominance 
$r_{\rm{Q}}$ remains approximately constant (see Fig.~\ref{A1_and_A2}) and inserting the solution \eqref{eq:DE-solution-Z} into the differential equation \eqref{eq: z_prime_matter}, we obtain the solution
\be
\mc{Q} = -\frac{4 k^2 s \lbrace 3 c_{A}^{2} \mathcal{R}_3 [ -\Phi_0 + r_{\rm{Q}} (\Phi_0 + \delta_{\mathrm{cdm}})] + \Phi_0 (-1 + 4 r_{\rm{Q}} + 3 w_{\rm{DE}})\rbrace}{3 c_{A}^{2} (9 \mathcal{H}^2+ 2 k^2 \mathcal{R}_3)} .
\label{eq:DE-solution-Q}
\ee
Fig.~\ref{Q_Z} depicts a comparison of our analytical approximate solutions for DE perturbations in the matter dominated regime [namely, Eqs.~\eqref{eq:DE-solution-Z}-\eqref{eq:DE-solution-Q}] against fully numerical solutions. We show a mode that remains smaller than the horizon while it evolves. While the agreement between analytical approximate and numerical solutions is quite good when matter dominates, as soon as DE becomes relevant our analytical solutions are no longer valid (see also lower panels in Fig.~\ref{Fig:perturbations-CLASS}).    

\section{Cosmological constraints: Methodology}
\label{section:cosmological-constraints}

Having discussed a few phenomenological aspects of the  coupled vector DE model in the previous sections, we now focus our attention on comparing the predictions of the model against available data sets. Our main goal is to compute cosmological constraints for parameters in the model. We also examine whether or not the coupled vector DE model is a plausible alternative for the standard $\Lambda$CDM in light of current tensions in parameters such as $H_0$ and $\sigma_8$.

In our analysis we use publicly available data sets including probes such as Supernova type Ia (\texttt{S}) from Ref.~\cite{Pan-STARRS1:2017jku}, Baryon Acoustic Oscillations (\texttt{B}) from Refs.~\cite{BOSS:2016wmc,2011,Ross:2014qpa}, and a Gaussian prior on the Hubble parameter $H_0$ (\texttt{H}) from Ref.~\cite{Riess:2021jrx} which mainly constrain the background evolution. As for data constraining linear order perturbations, we utilise CMB  temperature and polarisation (\texttt{P}), and CMB lensing (\texttt{L}) measurements provided by the Planck Collaboration from Ref.~\cite{Planck:2018vyg}.

As previously mentioned, in order to compute relevant observables of the cosmological model we use our modified version of the popular Boltzmann solver \texttt{CLASS}. In the code we implemented the system of differential equations governing the evolution of both background and linear order perturbations as discussed in Section~\ref{Sec: Theoretical framework and phenomenology}. The coupled vector DE cosmological model is described by $10$ parameters, namely: baryon density today $\omega_{\rm b}\equiv \Omega_{\rm{b}}h^2$; CDM density today $\omega_{\rm cdm}\equiv \Omega_{\rm{cdm}}h^2$; the Hubble expansion rate today $H_0$; log power of the primordial curvature perturbations $\ln 10^{10} A_{\rm s}$; scalar spectrum power-law index $n_{\rm s}$; Thomson scattering optical depth due to reionisation $\tau_{\rm reio}$; sum of neutrino masses $\sum m_\nu$; and the three specific parameters of the coupled vector DE model $s$, $p_2$, $Q$. Note that in our analysis we consider a single massive neutrino (i.e., two massless neutrinos). 

Since we aim at computing cosmological constraints, we carry out a statistical analysis by efficiently exploring the plausible parameter space of the coupled vector DE model. This is a pretty laborious task when dealing with a high dimensional parameter space: besides the cosmological parameters there are more than $20$ nuisance parameters taking into consideration other aspects such as systematic errors and astrophysical phenomena. We adapt the publicly available code \texttt{MontePython}~\cite{Audren:2012wb,Brinckmann:2018cvx}  in order to work along our modified version of \texttt{CLASS}. The statistical analysis is performed through a Markov Chain Monte Carlo (MCMC) technique~\cite{Lewis:2002ah,2017ARA&A..55..213S} by using the modified code \texttt{MontePython}. First, the latter picks up a random point in the allowed region of parameter space (see Table~\ref{tab:prior}) and calls \texttt{CLASS} which in turn solves the system of differential equations and computes relevant quantities such as the CMB angular power spectrum and luminosity distance. Second, the likelihood is evaluated. Third, since we do not have a proposal distribution for carrying out the sampling, we initially proceed with a Gaussian distribution having a diagonal covariance matrix. \texttt{MontePython} then explores the parameter space (i.e., comparing model predictions with observations via likelihood evaluations) eventually updating the proposal distribution by replacing the covariance matrix. The updating process goes on until the algorithm reaches an acceptance rate $\approx 0.25$ and the covariance matrix of the proposal distribution is frozen. Finally, \texttt{MontePython} performs the remaining part of parameter space exploration executing $\sim 10^5$ iterations. At this point of the analysis the Gelman-Rubin statistic for each varying parameter normally is  $\lesssim 10^{-2}$ and  convergence is attained. We can then compute all the relevant statistical information (e.g., 1D and 2D posteriors, mean values, confidence regions).
\begin{table}[http]
\centering 
\begin{tabular}{c c} 
\hline\hline 
Parameter & Prior range \\ 
\hline 
$s$ & [$0.024$, $1.$] \\
$p_2$ & [$0.024$, $1.$]\\
$Q$   &  [$-0.5$, $0.$]\\
\hline 
\end{tabular}
\caption{Flat prior bounds used in our statistical analysis. For other cosmological parameters we use the same prior range as specified in table 1 of Ref.~\cite{Planck:2013pxb}}
\label{tab:prior}
\end{table}

\section{Results}
\label{section:results}

\begin{figure*}[h]
\centering
\includegraphics[width=\textwidth]{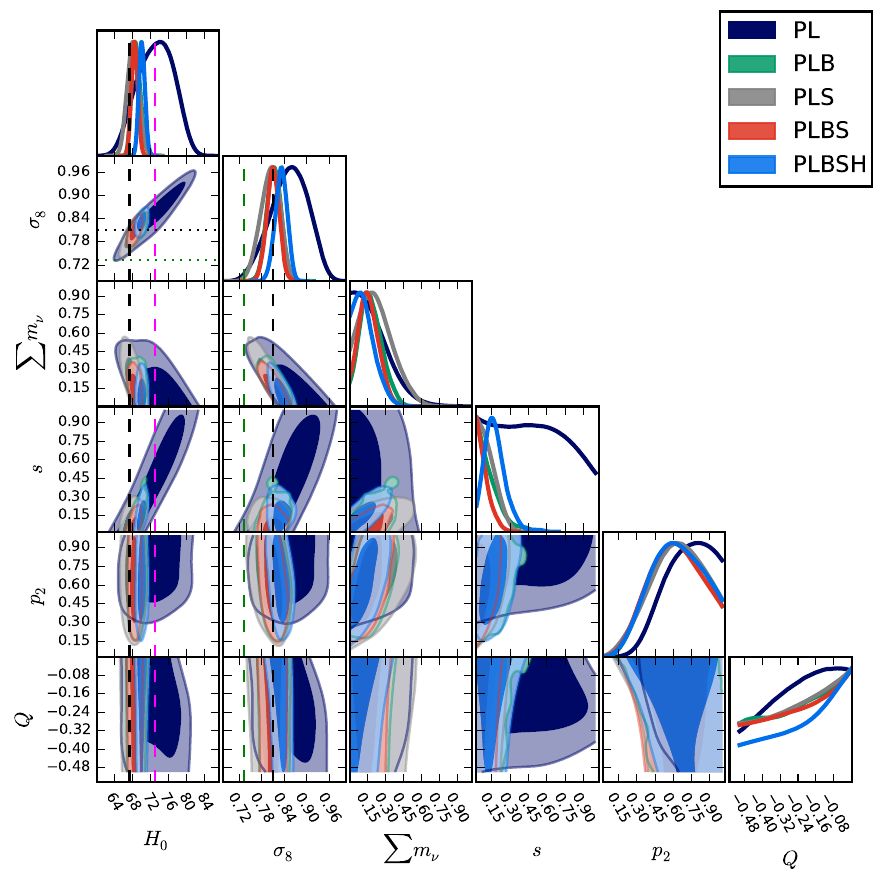}
\caption{1D marginalised likelihoods and confidence contours (i.e., $68\%$ and $95\%$) for parameters in the coupled vector DE model. \texttt{PL} stands CMB data including temperature, polarisation, and lensing; \texttt{B} Baryon Acoustic Oscillations; \texttt{S} supernova type Ia; \texttt{H} local measurement of the Hubble constant. Dashed, vertical and dotted, horizontal black lines indicate the $\Lambda$CDM baseline result reported by the Planck Collaboration (see column \texttt{TTTEEE+lowE+lensing} in table 2 of Ref.~\cite{Planck:2018vyg}). Vertical, dashed, magenta line indicates the SH0ES value $H_0=73.04\,\mathrm{km}\, \mathrm{s}^{-1}\,\mathrm{Mpc}^{-1}$ ~\cite{Riess:2021jrx}. Dashed, vertical and dotted, horizontal green lines indicate DES value $\sigma_8=0.733$ which uses three two-point correlation functions~\cite{DES:2021wwk}.}
\label{Fig:triangle-figure}
\end{figure*}

Here we present the results from our statistical analysis. Fig.~\ref{Fig:triangle-figure} depicts the 1D and 2D posterior distributions for the relevant cosmological parameters in the coupled vector DE model.\footnote{A full triangle plot for varying cosmological parameters in the coupled vector DE model can be found in Appendix~\ref{appendix:D}, see Fig.~\ref{Fig:triangle-figure-full}.} In Table~\ref{Table:constraints-background-perturbations} we summarise mean values and interval confidence when using different data sets. 

\begin{table*}[http]
\centering
\resizebox{\columnwidth}{!}{%
\begin{tabular}{c c c c c c}
\hline
Parameter  & \texttt{PL} & \texttt{PLB} & \texttt{PLS} & \texttt{PLBS} & \texttt{PLBSH} \\
\hline
$\omega_\mathrm{b} $ & $0.02241^{-0.00017}_{+0.00016}$ & $0.02242^{-0.00015}_{+0.00014}$ & $0.02238\pm 0.00017$ & $0.02244\pm 0.00014$ & $0.02246\pm 0.00014$\\
$\omega_{\mathrm{cdm}}$  & $0.1190^{-0.0014}_{+0.0013}$ & $0.1188\pm 0.0011$ & $0.1192\pm 0.0014$ & $0.1186\pm 0.0010$ & $0.1188^{-0.0011}_{+0.0012}$ \\
$H_0\, [\mathrm{km}\, \mathrm{s}^{-1}\,\mathrm{Mpc}^{-1}]$ &  $73.17^{-3.86}_{+4.32}$ & $68.86^{-1.09}_{+0.81} $ & $68.14^{-1.29}_{+1.28}$ & $68.48^{-0.74}_{+0.66}$ & $70.11^{-0.69}_{+0.64}$ \\
$ \sigma_8$  & $0.854^{-0.045}_{+0.055} $ & $0.813\pm 0.019$ & $0.802^{-0.023}_{0.029}$ & $0.810\pm 0.017$ & $0.832^{-0.015}_{+0.017}$\\
$n_{\rm{s}}$ &  $0.9657^{-0.0046}_{+0.0052}$ & $0.9665^{-0.0041}_{+0.0040}$ & $0.9650^{-0.0048}_{+0.0047}$ & $0.9672^{-0.0040}_{+0.0041}$ & $0.9670\pm 0.0041$ \\
$\tau_{\rm{reio}} $  & $0.0508^{-0.0080}_{+0.0077}$ & $0.0501^{-0.0076}_{+0.0080}$ & $0.0500\pm 0.0075$ & $0.0501\pm 0.0075$ & $0.0494^{-0.0071}_{+0.0083}$ \\
$\sum m_\nu\,[\mathrm{eV}] $  & $<0.246$ & $0.179^{-0.109}_{+0.076}$ & $0.226^{-0.179}_{+0.091}$ & $0.159^{-0.097}_{+0.071}$ & $ <0.157$ \\
$s$  & $<0.6380$ & $<0.157$ & $<0.148$ & $<0.104$ & $ 0.178^{-0.103}_{+0.058} $  \\
$p_2$ & $0.707^{-0.092}_{+0.282}$ & $0.624^{-0.190}_{+0.247}$ & $0.625^{-0.186}_{+0.259}$ & $0.611^{-0.220}_{+0.226}$ & $0.621^{-0.206}_{+0.239}$  \\
$Q$ & $>-0.292$ & $>-0.304$ & $>-0.3$ & $>-0.304$ & $>-0.273$ \\
\hline
\end{tabular}
}
\caption{Mean values and $68\%$ confidence limits on cosmological parameters for the coupled vector DE model.}
\label{Table:constraints-background-perturbations}
\end{table*}

Dashed, vertical and dotted, horizontal black lines in Fig.~\ref{Fig:triangle-figure} and Fig.~\ref{Fig:triangle-figure-full} indicate the $\Lambda$CDM baseline result reported by the Planck Collaboration. While the dashed, vertical, magenta line is the SH0ES value for $H_0$, dashed and dotted green lines indicate the DES value for $\sigma_8$. We can see that constraints for common parameters also shared by the coupled vector DE model are in good agreement with $\Lambda$CDM results. Regarding the parameters $s$, $p_2$, and $Q$, which only appear in the coupled vector DE model, we can notice that they are not fully constrained by the data sets utilised here.   

\section{Discussion}
\label{section:discussion}

Our analysis can be regarded as a generalisation of previous works (see, for instance,  Ref.~\cite{Heisenberg:2020xak}) since we allow more freedom in the parameter space: in addition to common parameters also appearing in the $\Lambda$CDM model, we marginalise over parameters $s$, $p_2$, and $Q$ characterising our DE model. Note that the coupled vector DE model was investigated in Ref.~\cite{Nakamura:2019phn}, but cosmological constraints were not computed and a detailed analysis of linear order perturbations also calculating observables was not carried out. 

To begin with,  we implemented the coupled vector DE model in the Boltzmann code \texttt{CLASS} as explained in previous sections. Our modified version of \texttt{CLASS} numerically solves the system of differential equations governing both background and linear order perturbations. An important detail of our implementation concerns setting up the initial conditions. Firstly, at the background level, we take advantage of the so-called `shooting method' implemented in \texttt{CLASS}. We use it to determine the DE parameter density $\Omega_{\rm{DE}}^0$ or equivalently the initial value of the vector field. In order to carry out this operation the code executes the background module a number of times starting from a `guess value' for the initial vector field.\footnote{Our `guess value' assumes that, during radiation dominance, the DE density behaves like $\rho_{\rm{DE}}=\rho_{\rm{DE}}^{\rm{ini}} a^{4s}$. The modified \texttt{CLASS} code we release further assumes that $\rho_{\rm{DE}}^{\rm{ini}}=10^{-x}$. Then, along with the cosmological parameters, the \texttt{CLASS} user must provide an starting $x$ value so that the `shooting method' finds a proper $x$ value satisfying the closure relation $\sum \Omega^0_{\rm{i}}=1$. We note that the `shooting method' along with our assumptions for $\rho_{\rm{DE}}$ during radiation dominance might fail depending on the region of parameter space: whereas for the prior ranges in Table~\ref{tab:prior} we were able to smoothly carry out our MCMC analyses, also allowing for positive values of the coupling $Q>0$ often led to failures in the `shooting method'. Since our statistical analysis relies on the ability for providing proper initial conditions so that \texttt{CLASS} can be executed $\approx 10^5$ times, in this work we opted for restricting ourselves to negative values of the coupling. Unless stated differently, we use $x=10$ as a default starting value in our computations.} In each iteration the code looks for a better initial vector field value so that the closure relation imposed by the Friedmann equation be satisfied. The process finishes when a given tolerance is reached. Secondly, regarding the initial conditions for the linear order perturbations, we keep the default behaviour in the code for CDM  perturbations while for DE perturbations we set $\mathcal{Q} = \mathcal{Z} = 0$. We also tried setting adiabatic initial conditions for DE perturbations, as explained in Appendix~\ref{appendix:B}, and found negligible changes in the solutions with respect to our choice $\mathcal{Q} = \mathcal{Z} = 0$.   

The dashed, vertical and dotted, horizontal black lines in Fig.~\ref{Fig:triangle-figure} and Fig.~\ref{Fig:triangle-figure-full} indicate the $\Lambda$CDM baseline result reported by the Planck Collaboration \cite{Planck:2018vyg}. We can see that when the coupled vector DE model is confronted with data, the common parameters  $\omega_{\rm{b}}$, $\omega_{\rm{cdm}}$, $\sigma_8$, $n_{\rm{s}}$, and $\tau_{\rm{reio}}$ are in good agreement with constraints for $\Lambda$CDM regardless of the data set utilised. However, the uncertainties are slightly increased due to a greater number of parameters in the model.  

Regarding the Hubble parameter $H_0$, we notice an increase in the uncertainty when using \texttt{PL} alone which artificially alleviates the tension with local measurements by virtue of changes in the angular diameter distance~\cite{Heisenberg:2020xak}. However, the inclusion of other data sets (i.e., \texttt{PLB}, \texttt{PLS}, \texttt{PLBS}) brings the value close to the Planck $\Lambda$CDM baseline result. When also taking into consideration the SH$0$ES prior (i.e., \texttt{PLBSH}) we found that our result sits in the middle (i.e., more than $2\sigma$ away) of the local measurement and the CMB $\Lambda$CDM inferred value  of $H_0$. We conclude that even though the coupled vector DE model might slightly alleviate the so-called Hubble tension, it does not provide a convincing solution.     

The results for the parameters $s$ and $p_2$ of the coupled vector DE model are in good agreement with findings for the uncoupled model investigated in Ref.~\cite{Heisenberg:2020xak}. In particular, while \texttt{PL} cannot constrain $s$, the inclusion of other data sets yields much better constraints. Although there is a preference for lower values of $s$, the constraining power of the data sets is not good enough and our results are prior dominated, namely, we are only able to put an upper bound on $s$. A similar situation occurs with the parameter $p_2$: in this case, however, our results are limited by the upper limit of the prior and we can only put a lower bound. Regarding the coupling parameter $Q$, we can see that even though values close to zero are favoured, it remains fully unconstrained for all data sets we utilise. 

Our investigation also confirms that current data sets are unable to measure the neutrino mass. We are only able to put upper bounds on $\sum m_\nu$. Note that previous works do not marginalise over the neutrino mass. Here we did include the neutrino mass in our analysis and found an interesting degeneracy $s-\sum m_\nu$. The inclusion of information from ongoing galaxy surveys like Euclid hence could improve the determination of $s$ as $\sum m_\nu$ is a target parameter in these kinds of experiments~\cite{PhysRevLett.95.011302}. Another relevant degeneracy when dealing with \texttt{PL} is $s-H_0$. The inclusion of supernovae and BAO, which constrain the background behaviour, significantly help in breaking the degeneracy.

The coupled vector DE model can make less negative the DE equation of state during matter dominance [see Eq.~\eqref{Eq:MD}] depending on the values of $s$ and $Q$~\cite{Nakamura:2019phn}. In particular, small values of $s$ and $Q<0$ (i.e., $r_{\rm{Q}}>0$) bring $w_{\rm{DE}}$ close to a cosmological constant (see Fig.~\ref{Fig:Background} and Fig.~\ref{A1_and_A2}).    

We worked out the behaviour of matter perturbations during matter dominance and found analytical approximate solutions [see Eqs.~\eqref{eq:CDM-solutions-theta-2}-\eqref{eq:CDM-solutions-delta-2}] that match those in the standard cosmological model. While our numerical solutions depicted in Fig.~\ref{delta_theta} indicate a relatively big impact of a non-vanishing coupling (see red and blue curves in the right-hand side panel) in the matter velocity perturbations, the matter density perturbations show a slight dependence on the coupling (see also Fig.~\ref{Fig:perturbations-CLASS}). Therefore we understand the good agreement on the strength of matter clustering $\sigma_8$ between the coupled vector DE model and $\Lambda$CDM (see Fig.~\ref{Fig:triangle-figure}).

\begin{figure*}[h]
\centering
\includegraphics[width=\textwidth]{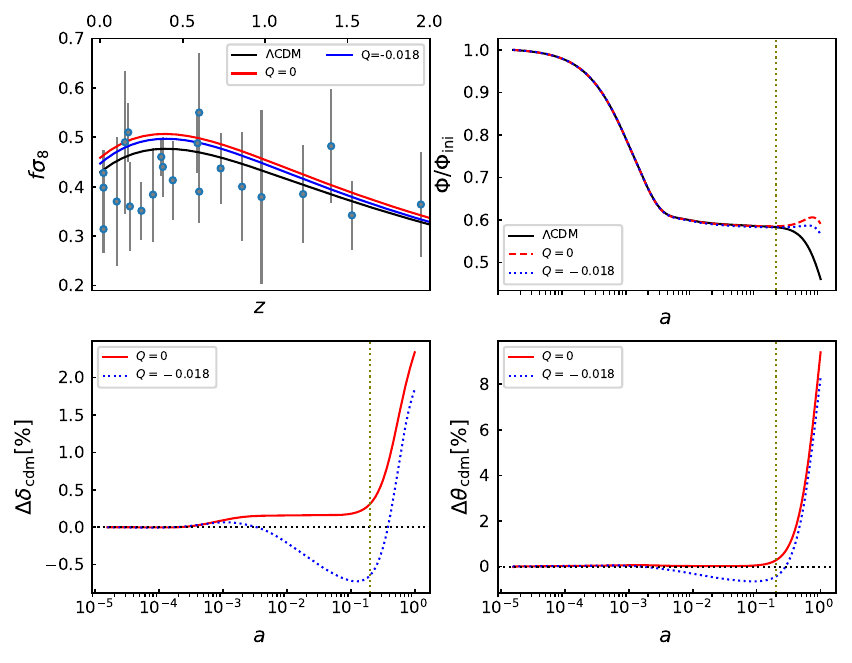}
\caption{Numerical solutions of linear order  perturbations in $\Lambda$CDM, uncoupled and coupled vector DE models as computed by our modified \texttt{CLASS} code. In the \textit{lower panels} the percentage difference relative to $\Lambda$CDM is depicted. \textit{Right upper panel} shows normalised gravitational potential $\Phi$. \textit{Upper left panel} shows the parameter $f\sigma_8$: whereas for the coupled (uncoupled) vector DE model $\sigma_{8,0}=0.828$ ($\sigma_{8,0}=0.848$), for $\Lambda$CDM $\sigma_{8,0}=0.811$. We use the best fit of our coupled vector DE model baseline result (i.e., the case using data set \texttt{PLBSH}): $\omega_{\rm{b}}=0.02235$, $\omega_{\rm{cdm}}=0.1196$, $\ln 10^{10} A_{\rm{s}}=3.035$, $n_{\rm{s}}=0.9645$, $H_0=69.86\,\rm{km}\,\rm{s}^{-1}\,\rm{Mpc}^{-1}$, $\tau_{\rm{reio}}=0.0523$, $\sum m_\nu = 0.16\,\rm{eV}$, $s=0.222$, $p_2=0.506$. Solutions for the uncoupled vector DE model were computed using (i.e., case \texttt{CMB+SNe+BAO+HST} in Ref.~\cite{Heisenberg:2020xak}): $\omega_{\rm{b}}=0.02248$, $\omega_{\rm{cdm}}=0.1195$, $\ln 10^{10} A_{\rm{s}}=3.040$, $n_{\rm{s}}=0.967$, $H_0=70.1\,\rm{km}\,\rm{s}^{-1}\,\rm{Mpc}^{-1}$, $\tau_{\rm{reio}}=0.0526$, $\sum m_\nu = 0.06\,\rm{eV}$, $s=0.1613$, $p_2=0.6887$.  $\Lambda$CDM solutions were computed using parameter values of the Planck baseline result. All panels  show the solutions for a mode $k=10^{-2}\,\rm{Mpc}^{-1}$. Vertical olive dotted lines approximately indicate the end of matter dominance.}
\label{Fig:fs8}
\end{figure*}

\begin{figure*}[h]
\centering
\includegraphics[width=\textwidth]{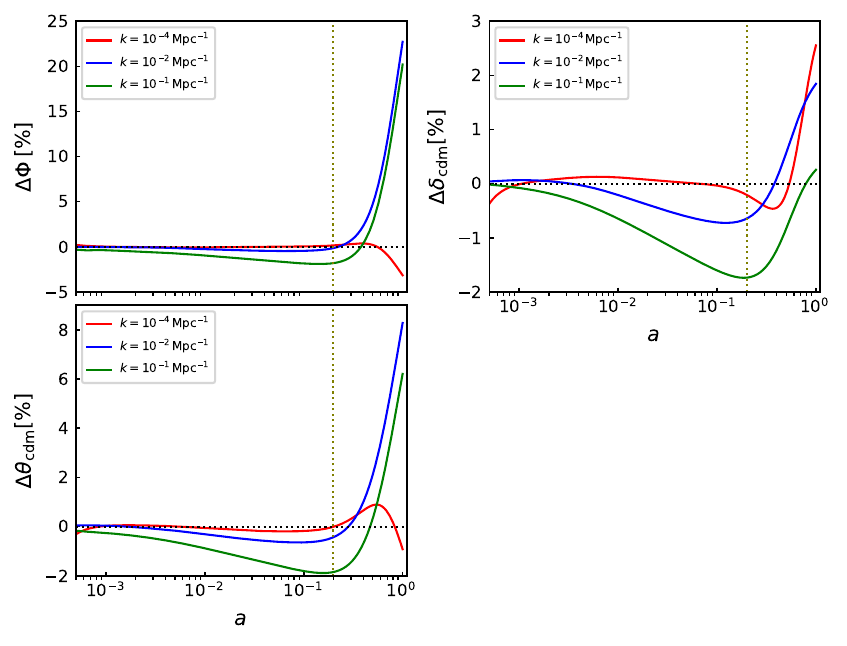}
\caption{Evolution of gravitational potential (\textit{upper left panel}), CDM density perturbation (\textit{upper right panel}), and CDM velocity perturbation (\textit{lower panel}). We depict the percentage difference of our best fit (see caption in Fig.~\ref{Fig:fs8}) baseline result \texttt{PLBSH} relative to the Planck Collaboration $\Lambda$CDM baseline result for three different wave numbers.}
\label{Fig:modes}
\end{figure*}

Information from Redshift Space Distortions (RSD) could further constrain the coupling parameter $Q$ due to its effect on the velocity perturbations. We use our modified \texttt{CLASS} code and numerically solve the full system of differential equations 
governing linear perturbations for a mode 
$k=10^{-2}\,\rm{Mpc}^{-1}$. 
We considered three cases: i) $\Lambda$CDM using the parameters values in the Planck baseline result~\cite{Planck:2018vyg}; ii) uncoupled vector DE model with parameters given tables 1 and 2 in Ref.~\cite{Heisenberg:2020xak} for the column \texttt{CMB+SNe+BAO+HST}; iii) coupled vector DE model with the best fit values of our baseline result \texttt{PLBSH}. The numerical solutions of the gravitational potential $\Phi$, and the CDM perturbations $\delta_{\rm{cdm}}$ and $\theta_{\rm{cdm}}$ are displayed in Fig.~\ref{Fig:fs8}. Having computed the numerical solutions for the matter perturbations, we can calculate the growth rate of matter perturbations function 
\ba 
f \sigma_8 (a) &\equiv& f(a) \, \sigma(a) \nn \\ &=&\sigma_{8,0}\, a \, \frac{\delta'_{\rm{cdm}}(a)}{\delta_{\rm{cdm}}(a=1)},
\label{Eq:fs8}
\ea
where $f(a)\equiv \dfrac{d\ln \delta_{\rm{cdm}}}{d\ln a}$ is the growth rate; $\sigma(a)\equiv \sigma_{8,0} \frac{\delta_{\rm{cdm}}(a)}{\delta_{\rm{cdm}}(a=1)}$ is the redshift-dependent rms fluctuations of the linear density field within spheres of radius $8\rm{Mpc}/h$, and $\sigma_{8,0}$ is its value today. Besides being a good discriminator of DE models, the function $f\sigma_8$ is independent of the bias in linear theory. In the left upper panel of Fig.~\ref{Fig:fs8} we show the $f\sigma_8$ results for the three cosmological models considered along with the RSD data compilation of Ref.~\cite{Sagredo:2018ahx}.  We can clearly see that our negatively coupled vector DE baseline result agrees better with the $\Lambda$CDM Planck baseline result than the uncoupled vector DE model. 
This is mainly caused by a different time evolution 
of the gravitational potential in the DE dominated epoch as well as the increase of matter velocity perturbations [see Eq.~\eqref{eq:perturbed_energy_for_cdm}] on small scales (see upper right panel in Fig.~\ref{Fig:fs8} and  Fig.~\ref{Fig:modes}). RSD measurements indicate a lower growth rate of matter perturbations than predicted by the standard cosmological model. Although the coupled vector DE model seemingly fits the RSD data better than the uncoupled vector DE model, it also moderately exacerbates the disagreement at low redshift. A positively coupled DE model can lead to a lower growth rate of matter perturbations than $\Lambda$CDM (see discussion in Appendix~\ref{appendix:E} and Fig.~\ref{Fig:fs8-Q-positive}).          

\section{Conclusions}
\label{section:conclusions}

Although the standard cosmological model stands as our current best description of the Universe, its phenomenological nature motivates the search for alternatives relying on more fundamental physics. Moreover, the quality of data at our disposal is starting to unveil interesting discrepancies in cosmological parameters when determined from different probes that enhance our curiosity in alternative scenarios to $\Lambda$CDM. 

In this work we explored the possibility of a vector field being the cause of the late-time Universe accelerating expansion. Since the nature of both CDM and DE remains unknown, we further considered an interaction in the dark sector. Previous works showed that a negative coupling could make the DE equation of state $w_{\rm{DE}}$ less negative during matter dominance, thus relieving the disagreement with observations~\cite{Nakamura:2019phn}. 

The coupled vector DE model that we investigated has three additional parameters with respect to $\Lambda$CDM. We carefully worked out the equations governing both background and linear perturbations. Due to our choice of the vector field configuration, which respects the observed isotropy and homogeneity on large scales, the coupling term does not add any additional contribution to the anisotropic stress. Therefore, during matter dominance and DE dominance we can consider the gravitational potentials $\Psi \approx \Phi$ in the Newtonian gauge. 

In order to better understand the dynamics of the coupled vector DE model, we looked for analytical approximate solutions of linear order perturbations. Since the vector field remains subdominant at the onset of evolution, we focused on a simplified scenario including only CDM and DE. Because the coupled vector DE model has a vanishing anisotropic stress, we were able to simplify the system of differential equations governing the linear order perturbations. We made a couple of assumptions regarding the evolution of the gravitational potential (i.e., quasi-static approximation) as well as scales deep into the horizon (i.e., sub-horizon approximation). We found analytical approximate solutions for both CDM and DE perturbations [see Eqs.~\eqref{eq:CDM-solutions-theta-2}-\eqref{eq:DE-solution-Q}]. We performed a comparison between numerical and analytical solutions and found a good agreement during the matter dominated regime.

The numerical solutions in the simplified scenario show slight differences in the growth of matter perturbations at late times with respect to $\Lambda$CDM. However, the negatively coupled vector DE model predicts bigger matter velocity perturbations than the standard model on small scales (see Fig.~\ref{delta_theta} and Fig.~\ref{Fig:modes}). Therefore RSD data help in further constraining a coupling in the dark sector. In fact, we found our negatively coupled vector DE is slightly disfavoured with respect to the standard cosmological model (see upper left panel in Fig.~\ref{Fig:fs8}). As we show in Appendix~\ref{appendix:E}, a positively coupled vector DE can lead to a lower growth rate of matter perturbations than predicted by $\Lambda$CDM, thus alleviating the disagreement with RSD at low red-shift (see Fig.~\ref{Fig:fs8-Q-positive}). While such a positively coupled vector DE may lead to a suppression of power on small scales in the matter power spectrum, it also may diminish the goodness of fit in the CMB angular power spectrum on small scales (see Fig.~\ref{Fig:CMBPk}).   

Our implementation of the full system of differential equations (i.e., background and linear order perturbations) in the Boltzmann code \texttt{CLASS} allowed us to compute predictions for observables such as the CMB angular power spectrum and the matter power spectrum (see Fig.~\ref{Fig:linear-perturbations}). We performed an statistical analysis including available data from the Cosmic Microwave Background anisotropies, Baryon Acoustic Oscillations, Supernovae type Ia as well as a Gaussian prior on the Hubble constant. Our cosmological constraints showed a preference for a vanishing coupling in the dark sector, but our result is dominated by the prior bounds. The other two parameters (i.e., $s$ and $p_2$) also present in the generalised Proca model~\cite{Heisenberg:2020xak} are also prior limited and we could only set an upper (lower) bound for $s$ ($p_2$) in good agreement with previous works.               

The investigation carried out in this work allows us to conclude that the negatively coupled vector DE cosmological model is a good fit for the available data utilised here. 
Our baseline result \texttt{PLBSH} (i.e., including CMB, BAO, supernovae, and $H_0$ data) is in good agreement with constraints found for common parameters in $\Lambda$CDM by the Planck Collaboration. Nevertheless, we found that our baseline result for the negatively coupled vector DE model is slightly  incompatible with RSD data due to a recent moderate growth of matter velocity perturbations and gravitational potential on small scales. Furthermore, since the coupled vector DE has three more parameters than $\Lambda$CDM, it is heavily disfavoured by Bayesian evidence. In addition, since the constraints for the Hubble constant $H_0$ and the strength of matter clustering $\sigma_8$ agree with Planck $\Lambda$CDM baseline results, the negatively coupled vector DE does not solve current tensions in $H_0$ and $\sigma_8$. A more sophisticated modelling (e.g., adding non-linear information) might be needed to clear up discrepancies in our understanding of current data~\cite{10.1093/mnras/stac2429,PhysRevLett.131.111001,ACT:2023dou,Knox:2019rjx}. 

\appendix

\section{Cosmological constraints}
\label{appendix:D}

\begin{figure*}[h]
\centering
\includegraphics[width=\textwidth]{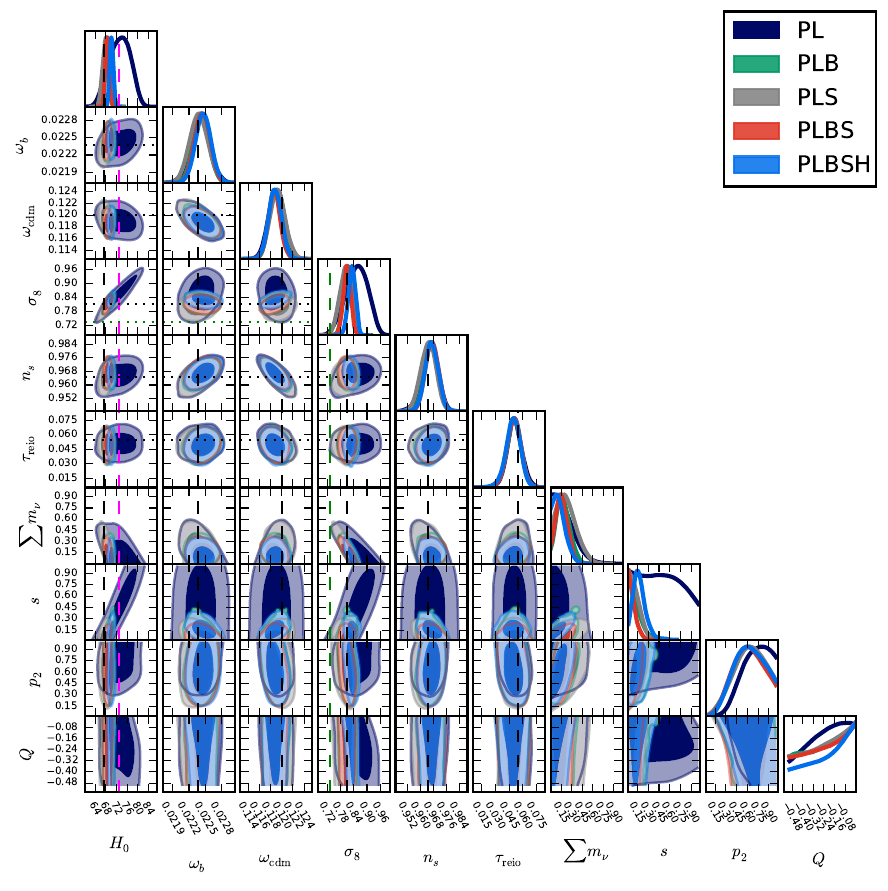}
\caption{1D marginalised likelihoods and confidence contours (i.e., $68\%$ and $95\%$) for parameters in the coupled vector DE model. \texttt{PL} stands CMB data including temperature, polarisation, and lensing; \texttt{B} Baryon Acoustic Oscillations; \texttt{S} supernova type Ia; \texttt{H} local measurement of the Hubble constant. Dashed, vertical and dotted, horizontal black lines indicate the $\Lambda$CDM baseline result reported by the Planck Collaboration (see column \texttt{TTTEEE+lowE+lensing} in table 2 of Ref.~\cite{Planck:2018vyg}). Vertical, dashed, magenta line indicates the SH0ES value $H_0=73.04\,\mathrm{km}\, \mathrm{s}^{-1}\,\mathrm{Mpc}^{-1}$ ~\cite{Riess:2021jrx}. Dashed, vertical and dotted, horizontal green lines indicate DES value $\sigma_8=0.733$ which uses three two-point correlation functions~\cite{DES:2021wwk}.}
\label{Fig:triangle-figure-full}
\end{figure*}

Figure~\ref{Fig:triangle-figure-full} shows 1D marginalised likelihoods and confidence contours for cosmological parameters in the negatively coupled vector DE model.

\section{A positively coupled vector DE}
\label{appendix:E}

As explained in Section~\ref{section:discussion} we could not fully explore the parameter space also including a positive coupling $Q>0$ in our analysis. An issue likely to be caused by our assumption of the behaviour of DE density during radiation dominance. Here however we present a few examples where the `shooting method' works properly, hence finding adequate initial conditions for the vector field and allowing the solution of the full system of differential equations governing background and linear order perturbations. For the current section we run our modified \texttt{CLASS} code using $x=7$ as a starting value for the `shooting method'. 

\begin{figure*}[h]
\centering
\includegraphics[width=\textwidth]{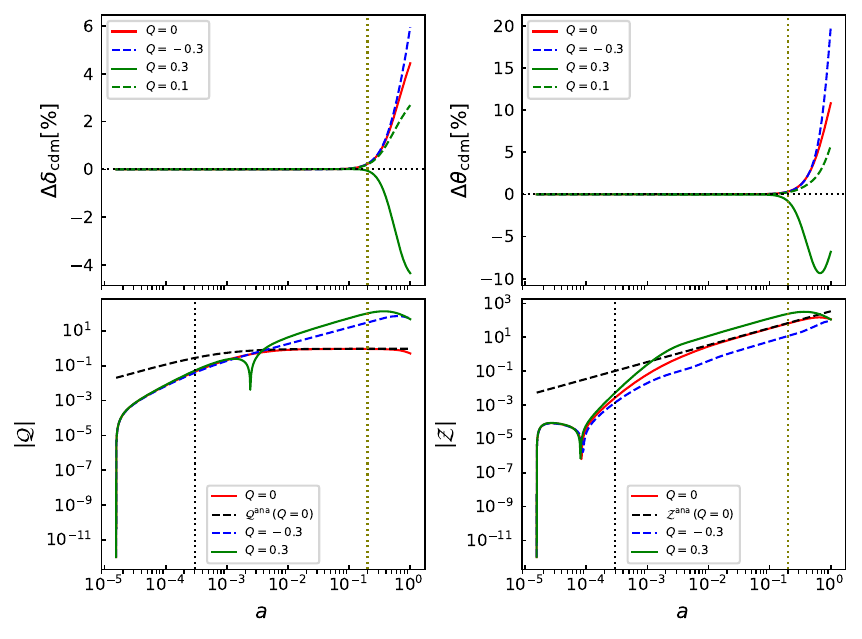}
\caption{\textit{Upper panels:} CDM density (left) and velocity (right) perturbations relative to the standard cosmological model. \textit{Lower panels:} variables related to DE perturbations. We show the output of our modified \texttt{CLASS} code for a mode $k=10^{-2}\,\mathrm{Mpc}^{-1}$.  As predicted by our analytical solutions, CDM perturbations in the coupled vector DE model closely follow $\Lambda$CDM solutions during during matter dominance. Common parameters to all models are: $\omega_{\rm{b}}=0.02235$, $\omega_{\rm{cdm}}=0.1196$, $\ln 10^{10} A_{\rm{s}}=3.035$, $n_{\rm{s}}=0.9645$, $H_0=69.86\,\rm{km}\,\rm{s}^{-1}\,\rm{Mpc}^{-1}$, $\tau_{\rm{reio}}=0.0523$, $\sum m_\nu = 0.16\,\rm{eV}$. Coupled vector DE model parameters are: $s=0.222$, $p_2=0.506$. Vertical dotted olive lines approximately indicate the end of matter dominance. Vertical dotted black lines approximately indicate horizon crossing for the depicted mode. Dashed black curves in lower panels show our approximate analytical solutions for sub-horizon scales during matter dominance using the value $\Phi_0=\Phi(a=0.1)$ from the numerical solution.}
\label{Fig:perturbations-CLASS}
\end{figure*}

In Fig.~\ref{Fig:perturbations-CLASS} we show solutions for CDM and DE perturbations in the coupled vector DE model and $\Lambda$CDM. As predicted by our analytical approximate solutions, CDM perturbations in the coupled vector DE model coincide during matter dominance with $\Lambda$CDM solutions (see also Fig.~\ref{delta_theta}). Lower panels in Fig.~\ref{Fig:perturbations-CLASS} depict solutions for DE perturbation variables in the coupled vector DE with a vertical black dotted line indicating horizon crossing. The evolution of $\mathcal{Q}$ and $\mathcal{Z}$ during matter dominance and after horizon crossing closely follows the behaviour described by our analytical approximate solutions (see also Fig.~\ref{Q_Z}).       

\begin{figure*}[h]
\centering
\includegraphics[width=\textwidth]{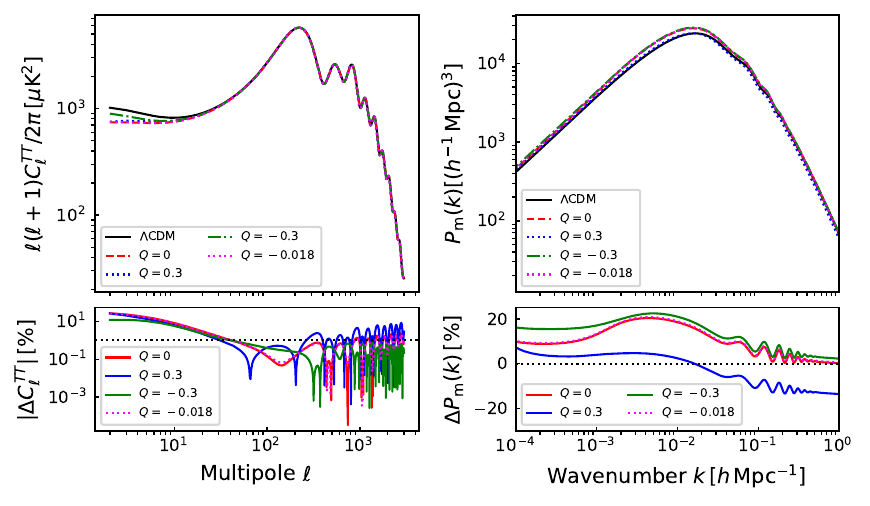}
\caption{\textit{Left:} CMB angular power spectrum and percentage difference relative to $\Lambda$CDM. \textit{Right:} Matter power spectrum and percentage difference relative to $\Lambda$CDM. Standard cosmological model results correspond to parameter values using the Planck baseline result. Parameter values for the coupled and uncoupled vector DE model are: $\omega_{\rm{b}}=0.02235$, $\omega_{\rm{cdm}}=0.1196$, $\ln 10^{10} A_{\rm{s}}=3.035$, $n_{\rm{s}}=0.9645$, $H_0=69.86\,\rm{km}\,\rm{s}^{-1}\,\rm{Mpc}^{-1}$, $\tau_{\rm{reio}}=0.0523$, $\sum m_\nu = 0.16\,\rm{eV}$, $s=0.222$, $p_2=0.506$. While our best fit has $Q=-0.018$, we also consider other values as shown in the legend.}
\label{Fig:CMBPk}
\end{figure*}

We now turn our attention to the possible effect a positive coupling could have on our baseline results. In Fig.~\ref{Fig:CMBPk} we compare results for the coupled vector DE model against $\Lambda$CDM Planck baseline. We compute CMB angular power spectrum and matter power spectrum for the best fit of our baseline result \texttt{PLBSH} while also considering different values of the coupling. 
Regarding CMB angular power spectrum (left panels), we see from the lower left panel that a more intense negative coupling leads to better agreement with $\Lambda$CDM on high and low multipole. An equally intense positive coupling worsen the agreement with $\Lambda$CDM particularly on small angular scales where error bars are smaller. Regarding the matter power spectrum (right panels), we note an overall unequal enhancement of power for a more intense negative coupling. The situation is different for an equally intense positive coupling as there is an enhancement (diminution) on large (small) scales. Such a behaviour could help in alleviating the disagreement on $\sigma_8$.        

\begin{figure*}[h]
\centering
\includegraphics[width=\textwidth]{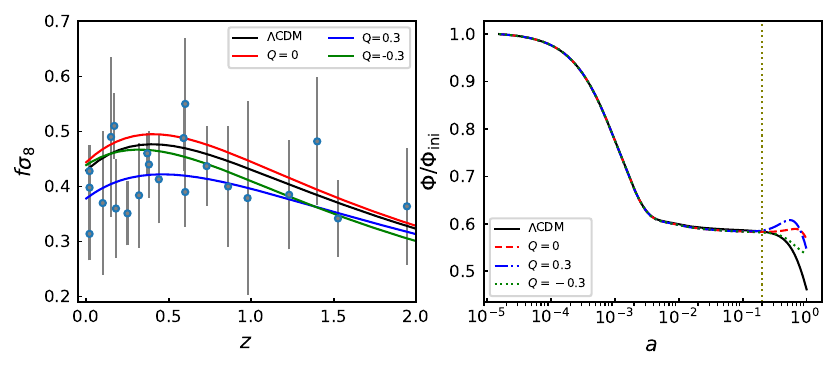}
\caption{\textit{Right panel} shows the normalised gravitational potential as computed from our modified \texttt{CLASS} code for a mode $k=10^{-2}\,\mathrm{Mpc}^{-1}$. \textit{Left panel} shows the parameter $f\sigma_8$. Cosmological parameters are as indicated in Fig.~\ref{Fig:CMBPk}.}
\label{Fig:fs8-Q-positive}
\end{figure*}

We computed $\sigma_{8,0}$ values in our examples of Fig.~\ref{Fig:CMBPk} and found: $\sigma_{8,0}=0.834$ for $Q=-0.3$; $\sigma_{8,0}=0.827$ for $Q=0$; $\sigma_{8,0}=0.767$ for $Q=0.3$. Then a positively coupled vector DE model could help in alleviating the so called $\sigma_8$ tension. In Fig.~\ref{Fig:fs8-Q-positive} we display changes in the growth rate of matter perturbations introduced by more intense couplings than our best fit value $Q=-0.018$. Due to recent changes in the evolution of the gravitational potential and an enhancement CDM velocity perturbations on small scales, a positively coupled vector DE model might play a part in lowering the growth rate of matter perturbation with respect to $\Lambda$CDM. A detailed statistical analysis also allowing $Q>0$ needs to be carried out as the CMB data constraining power on small scales is high and therefore small deviations from the standard cosmological model $C_\ell^{TT}$ on high $\ell$ are expected.     

\section{Detailed equations}
\label{appendix:A}
We start with the action
\be
\mathcal{S} =\int d^{4}x\sqrt{-g}\left[\frac{M_{\mathrm{pl}}^{2}}{2}R+ \mathcal{L}_{2} + \mathcal{L}_{3} +  \mathcal{L}_{\mathrm{int}} + \mathcal{L}_{\mathrm{r}}+ \mathcal{L}_{\mathrm{b}} + \mathcal{L}_{\mathrm{cdm}} \right],
\label{Eq:action}
\ee
where
\be
\mathcal{L}_{2} = G_{2}(X,F), \qquad \mathcal{L}_{3} = G_{3}(X), \qquad X = -\frac{1}{2}A_{\mu}A^{\mu}, \qquad F = -\frac{1}{4}F_{\mu\nu}F^{\mu\nu},
\ee
and
\be
\mathcal{L}_{\mathrm{int}} = Q f(X)\mathcal{L}_{\mathrm{cdm}}.
\ee
Varying Eq.~\eqref{Eq:action} with respect to the metric tensor
\be
G_{\mu\nu} =   T^{\mathrm{r}}_{\mu\nu}+ T^{\mathrm{b}}_{\mu\nu} + T_{\mu\nu},
\label{eq:1}
\ee
where the energy momentum tensor $T_{\mu\nu}$ denotes the sum of contributions from both DE and effective dark matter
\ba 
T_{\mu\nu} &\equiv& T^{\mathrm{DE}}_{\mu\nu} + \tilde{T}^{\mathrm{cdm}}_{\mu\nu} = G_{2F} F_{\mu}^{\alpha}F_{\nu\alpha}  + 
G_{2}g_{\mu\nu} + G_{2X}A_{\mu}A_{\nu} \nn \\ 
&+& G_{3X}\[A_{\mu}A_{\nu}\nabla_{\alpha}A^{\alpha}+g_{\mu\nu}A^{\alpha}A^{\beta}\nabla_{\beta}A_{\alpha}-A^{\alpha}\(A_{\nu}\nabla_{\mu}A_{\alpha}+A_{\mu}\nabla_{\nu}A_{\alpha}\)\]  \nn \\
&+&  Qf_{X}A_{\mu}A_{\nu}\mathcal{L}_{\mathrm{cdm}} +(1+Qf(X))T^{\mathrm{cdm}}_{\mu\nu}.
\ea
The vector field equation is obtained varying the action Eq.~\eqref{Eq:action} with respect to $A_{\mu}$ 
\ba
\epsilon^{\mu} \equiv  & G_{2F} \nabla_\nu F^{\nu\mu} + G_{2XF}A^\nu \nabla_\alpha A_\nu F^{\mu \alpha} + G_{2FF}\nabla^\mu A^\nu \nabla^\beta A^\alpha \nabla_\nu F_{\beta\alpha} -Qf_{X}\mathcal{L}_{\mathrm{cdm}}A^{\mu} \nn  \\
& + G_{2FF}\nabla^\nu A^\mu \nabla^\beta A^\alpha \nabla_\nu F_{\alpha\beta} - G_{2X}A^{\mu}-G_{3X}(A^{\mu}\nabla_{\alpha}A^{\alpha}-A^{\alpha}\nabla^{\mu}A_{\alpha}) = 0. 
\ea
Here we consider $G_{2}$ as a linear function of $F$, then $G_{2F}$ is a constant and $G_{2XF} = G_{2FF} = 0$. As a result $\epsilon^{\mu}$ simplifies and reads  
\be
\epsilon^{\mu} \equiv  G_{2F} \nabla_\nu F^{\nu\mu}  - G_{2X}A^{\mu}-G_{3X}(A^{\mu}\nabla_{\alpha}A^{\alpha}-A^{\alpha}\nabla^{\mu}A_{\alpha}) - Qf_{X}\mathcal{L}_{\mathrm{cdm}}A^{\mu} = 0. 
\label{eq:2}
\ee 
The condition of energy-momentum conservation (i.e., $\nabla_{\mu}T^{\mu}_{\nu} = 0$) is satisfied if  
\be
\nabla_{\mu}\[(1+Qf(X))T^{(\mathrm{cdm})~\mu}_{~\nu} \] + Qf_{X}\mathcal{L}_{\mathrm{cdm}}A^{\mu}\nabla_{\nu}A_{\mu} = 0.
 \label{eq:3}
\ee
\subsection{Background evolution}
Our vector field configuration is
\be
A^{\mu} = \( \frac{\phi}{a},0,0,0\).
\ee
From Eq.~\eqref{eq:1} we obtain the Friedmann equations
\be
a^{-2}\mathcal{H}^{2} = \rho_{\rm{r}}+\rho_{\rm{b}} + \left[1+Qf(X)\right]\rho_{\rm{cdm}} -G_{2}(X),
\ee
\be
a^{-2}(2\mathcal{H}'+\mathcal{H}^{2})= -P_{\rm{r}} -G_{2}(X)+a^{-1}\phi^{2}\phi'G_{3,X}(X).    
\ee
From Eq.~\eqref{eq:2} we obtain the vector field equation of motion
\be
G_{2X}(X)+3a^{-1}\mathcal{H}\phi G_{3X}(X)-Qf_{X}(X)\rho_{\mathrm{cdm}} = 0.    
\label{eq:4}
\ee
We defined $\rho_{\mathrm{DE}}=-G_{2}$, and $P_{\mathrm{DE}} = G_{2}(X) - a^{-1}\phi^{2}\phi'G_{3X}$.
The continuity equation is obtained by taking the component $\nu= 0$ in  Eq.~\eqref{eq:3} 
\be
\partial_{\tau}\left\{ \left[1+Qf(X)\right]\rho_{\mathrm{cdm}}\right\} + 3\mathcal{H}\[(1+Qf(X)\]\rho_{\mathrm{cdm}} = \frac{\phi\phi' Qf_{X}}{1+Qf(X)}\[1+Qf(X)\]\rho_{\mathrm{cdm}}.
\ee
In addition, taking the derivative of $G_{2}(X)$ with respect to the conformal time and using Eq.~\eqref{eq:4}, we find  
\be
\rho_{\mathrm{DE}}' + 3\mathcal{H}\( \rho_{\mathrm{DE}} + P_{\mathrm{DE}} \) = -\frac{\phi\phi' Qf_{X}}{1+Qf(X)}\[1+Qf(X)\]\rho_{\mathrm{cdm}}.
\label{eq:15}
\ee

\subsection{Linear order perturbations}
The field configuration is
\be
\delta A^{\mu} = \(\frac{\delta\phi}{a},\vec{\nabla} \frac{\chi}{a^{2}}\).
\ee
From Eq.~\eqref{eq:2} we obtain the perturbed vector field equations of motion 
\ba
\delta\epsilon_{0} &=& k^{2}a^{-2}G_{2F}\[a\(\delta\phi+2\phi\Psi\)+\chi' \] \nn \\ 
&+& \[a\phi^{2}G_{2XX} +3\mathcal{H}\phi\(\phi^{2}G_{3XX}+G_{3X}\)-a\phi^{2}Qf_{XX}\rho_{\rm{cdm}}\]\(\delta\phi+\phi\Psi\) \nn \\
&-&k^{2}a^{-1}\phi G_{3X}\chi-a\phi Qf_{X}\delta\rho_{\mathrm{cdm}}-3\phi^{2}G_{3X}\(\mathcal{H}\Psi+\Phi'\) =0,
\label{eq:5}
\ea
\ba
i k_{j}\delta\epsilon_{j} &=& k^{2}a^{-2}G_{2F}\{\partial_{\tau}\[a\(\delta\phi+2\phi\Psi\)\]+\chi'' \} \nn \\
&+& k^{2}\phi G_{3X}\(\delta\phi+\phi\Psi\)+k^{2}a^{-1}\phi'G_{3X}\chi = 0. 
\label{eq:6}
\ea
The linear perturbations of the gravitational field equations are given by 
\ba
2k^{2}\Phi &+& 6\mathcal{H}\(\mathcal{H}\Psi+\Phi'\) = 8\pi G  a^{2}\left\{ \delta\rho_{\mathrm{r}} +\delta\rho_{\mathrm{b}} +\[1+Qf(X)\]\delta\rho_{\mathrm{cdm}} -\phi^{2}Qf_{X}\delta\rho_{\mathrm{cdm}} \right.  \nn \\
&+& \[\phi^{2}G_{2XX} +3a^{-1}\mathcal{H}\phi\(\phi^{2}G_{3XX}+2G_{3X}\)-\phi^{2}Qf_{XX}\rho_{\rm{cdm}}\]\(\phi\delta\phi+\phi^{2}\Psi\)  \nn \\
&-& \left.  k^{2}a^{-2}\phi^{2}G_{3X}\chi -3 a^{-1}\phi^{3}G_{3X}\(\mathcal{H}\Psi+\Phi' \) \right\},
\label{eq:11}
\ea 
\ba
2k^{2}\(\mathcal{H}\Psi +\Phi' \) &=& 8\pi G a^{2}\left\{\(\rho_{\mathrm{r}} + P_{\mathrm{r}}\)\theta_{\mathrm{r}} + \rho_{\mathrm{b}}\theta_{\mathrm{b}} + \[1+Qf(X)\]\rho_{\mathrm{cdm}}\theta_{\mathrm{cdm}}   \right. \nn \\
&-& \left. k^{2}a^{-1}\phi^{2}G_{3X}\(\delta\phi+\phi\Psi\)\right\}, 
\label{eq:12}
\ea

\ba
&2\Phi''& + 2\mathcal{H}\( \Psi' + 2\Phi'\)+2\(\mathcal{H}^{2}+2\mathcal{H}'\)\Psi + \frac{2}{3}k^{2}\(\Phi-\Psi\) =  8\pi G a^{2} \{\delta P_{\mathrm{r}}\nn \\\
&+&\phi G_{2X}\(\delta\phi+\phi\Psi\)-a^{-1}\partial_{\tau}\[\phi^{2}G_{3X}\(\delta\phi+\phi\Psi\)\] + a^{-1}\phi^{2}\phi'G_{3X}\Psi\}. 
\label{eq:13}
\ea

\subsection{Evolution of vector field perturbations}

The free functions of the coupled vector DE model are polynomials
\be
G_{2}(X) = b_{2}X^{p_{2}}, \qquad G_{3} = b_{3}X^{p_{3}}, \qquad f(X) = \hat{X}^{q}, \qquad \mathrm{where} \qquad \hat{X} \equiv \frac{X}{X_{0}}.
\ee
Then 
\be
G_{2X} =\frac{p_{2}G_{2}}{X}, \qquad G_{3X} = \frac{p_{3}G_{3}}{X}, \qquad f_{X} = \frac{q f(X)}{X},
\ee
\be
G_{2XX} =p_{2}(p_{2}-1)\frac{G_{2}}{X^{2}}, \qquad G_{3XX} = p_{3}(p_{3}-1)\frac{G_{3}}{X^{2}}, \qquad f_{XX} = q(q-1)\frac{ f(X)}{X^{2}}.
\ee
From Eq.~\eqref{eq:4} $G_{3X}$ can be written in terms of $G_{2X}$ and $f(X)$ as
\ba
G_{3X} &=& \frac{1}{3}\frac{a}{\mathcal{H}\phi}G_{2X}\(\frac{Qf_{X}\rho_{\mathrm{cdm}}}{G_{2X}}-1 \)  \nonumber \\
 &=& -\frac{1}{3}\frac{a}{\mathcal{H}\phi}G_{2X}(1-r_{\rm{Q}}),
\ea
where we define
\be
r_{\rm{Q}} \equiv \frac{Qf_{X}\rho_{\mathrm{cdm}}}{G_{2X}}.
\label{Eq:rQ-definition}
\ee
We can write the derivative of Eq.~\eqref{Eq:rQ-definition} as
\be
r_{\rm{Q}}' = \mathcal{H}r_{\rm{Q}}\[\frac{q-p_{2}}{p_{2}}\epsilon_{\phi}-3\],  \qquad \epsilon_{\phi}\equiv \frac{\rho_{\mathrm{DE}}'}{\mathcal{H}\rho_{\mathrm{DE}}} = \frac{\phi'}{2p_{2}\mathcal{H}\phi}.
\ee
It is helpful to define the following auxiliary variables  
\ba
\delta_{\phi} &\equiv& \frac{k\(\delta\phi+\phi\Psi\)}{\phi'},\label{Eq:delta-phi}\\
\delta_{\chi} &\equiv& \frac{k\chi}{a\phi}, \label{Eq:delta-chi-definition}\\
\mathcal{Z} &\equiv& -G_{2F}\frac{k^{2}\phi}{a^{3}\rho_{\rm{DE}}}\[a\(\delta\phi+2\phi\Psi\) +\chi'\] \label{Eq:Z-definition}.
\ea
Using the variables \eqref{Eq:delta-phi}-\eqref{Eq:Z-definition}, Eq.~\eqref{eq:5} can be rewritten as
\be
-\mathcal{Z} + p\(1+s_{1}r_{\rm{Q}}\) \frac{\mathcal{H}}{k}\epsilon_{\phi}\delta_{\phi}-\frac{2}{3}\frac{k}{\mathcal{H}}p_{2}(1-r_{\rm{Q}})\delta_{\chi} + 2p_{2}r_{\rm{Q}}\delta_{\mathrm{cdm}}-\frac{2p_{2}}{\mathcal{H}}(1-r_{\rm{Q}})(\mathcal{H}\Psi+\Phi') = 0,
\label{Eq:perturbed-vector-field-0}
\ee
where
\be
p = 2p_{3}-2p_{2}-1, \qquad s_{1}=\frac{2q-2p_{3}-1}{p}, \qquad \delta_{\mathrm{cdm}}= \frac{\delta\rho_{\mathrm{cdm}}}{\rho_{\mathrm{cdm}}}. 
\ee
Then from Eq.~\eqref{Eq:perturbed-vector-field-0}
\be
\epsilon_{\phi}\delta_{\phi} = \frac{1}{1+s_{1}r_{\rm{Q}}}\left\{\frac{k}{p\mathcal{H}}\mathcal{Z} +\frac{2}{3}\frac{s k^{2}}{H^{2}}(1-r_{\rm{Q}})\delta_{\chi} + \frac{2sk}{\mathcal{H}^{2}}(1-r_{\rm{Q}})(\mathcal{H}\Psi+\Phi') - \frac{2ks}{\mathcal{H}}r_{\rm{Q}}\delta_{\mathrm{cdm}}\right\},
\label{eq:7}
\ee
with $s=p_{2}/p$. The spatial component of the perturbed vector field equation Eq.~\eqref{eq:6} can also be rewritten using the variables \eqref{Eq:delta-phi}-\eqref{Eq:Z-definition}, it reads
\be
-\mathcal{Z}' +\mathcal{H}\mathcal{Z}\[\epsilon_{\phi}\(\frac{1}{2p_{2}}-1\)-3 \]+\frac{1}{2}\frac{k}{\mathcal{H}}(1-r_{\rm{Q}})\epsilon_{\phi}\(\delta_{\phi}+\delta_{\chi}\) = 0.
\label{eq:8}
\ee
Using the definitions \eqref{Eq:delta-phi}-\eqref{Eq:Z-definition} we can calculate $\delta_{\chi}'$
\be
\delta_{\chi}' = -\(\frac{\epsilon_{\phi}}{2p_{2}}+1\)\mathcal{H}\delta_{\chi}-\frac{\epsilon_{\phi}}{2p_{2}}\mathcal{H}\delta_{\phi}-\frac{a^{2}\rho_{\mathrm{DE}}}{G_{2F}k\phi^{2}}\mathcal{Z}-k\Psi.
\label{eq:9}
\ee
We can now introduce a new variable
\be
\mathcal{Q} \equiv -\(\frac{\mathcal{Z}}{p}+\frac{2}{3}\frac{s k}{\mathcal{H}}(1-r_{\rm{Q}})\delta_{\chi}\).
\label{eq:10}
\ee
If we utilise Eqs.~\eqref{eq:7}-\eqref{eq:10}, we can find the derivative in terms of $\mathcal{Z}$ and $\mathcal{Q}$
\be
\mathcal{Q}'  = -2\(1-\frac{3}{4}\alpha \)\mathcal{H}\mathcal{Q}+\(s\mathcal{R}_{1}+3\(1+s\mathcal{R}_{1})\)c_{A}^{2}\)\mathcal{H}\mathcal{Z} + \frac{2}{3}\frac{k^{2}s}{\mathcal{H}}(1-r_{\rm{Q}})\Psi,
\ee
where
\be
c_{A}^{2}=p^{-1}\(1+s\mathcal{R}_{1}\)^{-1}\left[\frac{2sp}{3\phi^{2}_{\rm{pl}}}\mathcal{R}_{1} + \frac{1}{3}\(1-sp\mathcal{R}_{1}\)+\frac{1}{2}\(1+2s-\frac{1}{p}\)\alpha \right], 
\ee
\be
\mathcal{R}_{1} = (1-r_{\rm{Q}})\Omega_{\mathrm{DE}}, \qquad \phi_{\rm{pl}} = \frac{\phi}{M_{\mathrm{pl}}}.
\ee
We can also rewrite Eq.~\eqref{eq:8} in terms of the $\mathcal{Q}$ by using Eqs.~\eqref{eq:7}, \eqref{eq:9}-\eqref{eq:10}  
\ba
\mathcal{Z}' &=& 3\frac{r_{\rm{Q}}+w_{\mathrm{DE}}}{1-r_{\rm{Q}}}\mathcal{H}\mathcal{Z}-\(\frac{k^{2}}{3\mathcal{H}^{2}}\frac{1-r_{\rm{Q}}}{1+s_{1}r_{\rm{Q}}} + \frac{3}{2}\alpha\)\mathcal{Q}+ \frac{2}{3}\frac{k^{2}s}{\mathcal{H}^{2}}\frac{(1-r_{\rm{Q}})^{2}}{1+s_{1}r_{\rm{Q}}}(\mathcal{H}\Psi+\Phi') \nn \\
&-& \frac{2}{3}\frac{k^{2}s}{\mathcal{H}}\frac{1-r_{\rm{Q}}}{1+s_{1}r_{\rm{Q}}}r_{\rm{Q}}\delta_{\mathrm{cdm}}.
\ea

\subsection{Energy continuity and Euler equations}
\label{appendix: Euler_equations}

Taking into consideration the component $\nu=0$ of Eq.~\eqref{eq:3} and focusing on the linear perturbation, we get
\ba
& &\partial_{\tau}\left\{\left[1+Qf(X)\right]\delta\rho_{\mathrm{cdm}}\right\} + 3\mathcal{H}\left[1+Qf(X)\right]\delta\rho_{\mathrm{cdm}}+\left[1+Qf(X)\right]\rho_{\mathrm{cdm}}\theta_{\mathrm{cdm}} \nn \\
&-& 3\left[1+Qf(X)\right]\rho_{\mathrm{cdm}}\Phi'  = \phi\phi'Qf_{X}\delta\rho_{\mathrm{cdm}}.
\label{Eq:perturbed-energy-momentum-0}
\ea
The component $\nu = i$ of Eq.~\eqref{eq:3}  is computed likewise, it yields 
\ba
&&\partial_{\tau}\left\{\left[1+Qf(X)\right]\theta_{\mathrm{cdm}}\right\} + 4\mathcal{H}\left\{\left[1+Qf(X)\right]\theta_{\mathrm{cdm}}\right\} - \mathcal{H}^{2}\left[1+Qf(X)\right]\rho_{\mathrm{cdm}}\Psi \nn \\
&& =k^{2}Qf_{X}\rho_{\mathrm{cdm}}\(\phi\delta\phi+\phi^{2}\Psi\).
\label{Eq:perturbed-energy-momentum-i}
\ea 
Eqs.~\eqref{Eq:perturbed-energy-momentum-0}-\eqref{Eq:perturbed-energy-momentum-i} motivate us to define effective CDM perturbations
\be
\delta\tilde{\rho}_{\mathrm{cdm}} \equiv \left[1+Qf(X)\right]\delta\rho_{\mathrm{cdm}}, \qquad \tilde{\rho}_{\mathrm{cdm}}\tilde{\theta}_{\mathrm{cdm}} \equiv \left[1+Qf(X)\right]\rho_{\mathrm{cdm}}\theta_{\mathrm{cdm}}.
\ee
From Eqs.~\eqref{eq:11}-\eqref{eq:13} we define the effective DE perturbations 
\ba
\delta\rho_{\mathrm{DE}} &\equiv& \[\phi^{2}G_{2XX} +3a^{-1}\mathcal{H}\phi\(\phi^{2}G_{3XX}+2G_{3X}\)-\phi^{2}Qf_{XX}\rho_{c}\]\(\phi\delta\phi+\phi^{2}\Psi\) \nn \\
&-&  k^{2}a^{-2}\phi^{2}G_{3X}\chi -3 a^{-1}\phi^{3}G_{3X}\(\mathcal{H}\Psi+\Phi' \) -\phi^{2}Qf_{X}\delta\rho_{\mathrm{cdm}},
\label{Eq:DE-density-perturbation}
\ea
\be
\(\rho_{\mathrm{DE}}+P_{\mathrm{DE}}\)\theta_{\mathrm{DE}} \equiv  -k^{2}a^{-1}\phi^{2}G_{3X}\(\delta\phi+\phi\Psi\),
\label{Eq:DE-velocity-perturbation}
\ee
\be
\delta P_{\mathrm{DE}} \equiv \phi G_{2X}\(\delta\phi+\phi\Psi\)-a^{-1}\partial_{\tau}\[\phi^{2}G_{3X}\(\delta\phi+\phi\Psi\)\] + a^{-1}\phi^{2}\phi'G_{3X}\Psi.
\label{Eq:DE-pressure-perturbation}
\ee
Note that deriving Eq.~\eqref{Eq:DE-velocity-perturbation}  we obtain the perturbed Euler equation for DE
\ba
\partial_{\tau}\left[\(\rho_{\mathrm{DE}}+P_{\mathrm{DE}}\)\theta_{\mathrm{DE}} \right] &+& 4\mathcal{H}\(\rho_{\mathrm{DE}}+P_{\mathrm{DE}}\)\theta_{\mathrm{DE}} -k^{2}\delta P_{\mathrm{DE}} -k^{2}\(\rho_{\mathrm{DE}}+P_{\mathrm{DE}}\)\Psi = \nn \\
&-& k^{2}Qf_{X}\rho_{\mathrm{cdm}}\(\phi\delta\phi+\phi^{2}\Psi\).
\label{eq:16}
\ea
In terms of $\mathcal{Z}$ and the DE perturbations \eqref{Eq:DE-density-perturbation}-\eqref{Eq:DE-pressure-perturbation}, the perturbed vector field equation \eqref{eq:5} reads 
\be
-\rho_{\mathrm{DE}} \mathcal{Z} + \delta\rho_{\mathrm{DE}} + \frac{3\mathcal{H}}{k^{2}}\(\rho_{\mathrm{DE}} + P_{\mathrm{DE}}\)\theta_{\mathrm{DE}} = 0.
\label{eq:14}
\ee
If we derive Eq.~\eqref{eq:14} and utilise Eqs.~\eqref{eq:15}, \eqref{eq:8}, \eqref{eq:10}, and \eqref{eq:16}, we find the differential equation governing the DE density perturbation 
\ba
\delta\rho_{\mathrm{DE}}' + 3\mathcal{H}\(\delta\rho_{\mathrm{DE}}+\delta P_{\mathrm{DE}}\) &+& \(\rho_{\mathrm{DE}}+P_{\mathrm{DE}}\)\theta_{\mathrm{DE}} -3\(\rho_{\mathrm{DE}}+P_{\mathrm{DE}}\)\Phi' = \nn \\
&-&\phi\phi'Qf_{X}\delta\rho_{\mathrm{cdm}}.
\ea

\section{Adiabatic initial conditions}
\label{appendix:B}

We use 
\be
\frac{\delta_{\rm{DE}}}{1+w_{\rm{DE}}} - \delta_{\rm{cdm}} = 0,
\label{Eq: adiabatic_initial_condition}
\ee
as a starting point to derive the adiabatic initial conditions for dark energy perturbations during radiation dominance and under the super-horizon approximation. Since we are dealing with two variables describing DE perturbations (i.e., $\mc{Q}$ and $\mc{Z}$), an additional equation is required to determine the initial conditions. We proceed by considering the derivative of Eq.~\eqref{Eq: adiabatic_initial_condition}
\be
\(\frac{\delta_{\rm{DE}}}{1+w_{\rm{DE}}}\)'- \delta_{\rm{cdm}}' = 0.
\label{Eq: adiabatic_initial_condition_2}
\ee
The DE density perturbation can be written as 
\be
\delta_{\rm{DE}} = \mc{Z}-\frac{3\cH}{k^{2}}\(1+w_{\rm{DE}}\)\theta_{\rm{DE}}.
\label{Eq: delta_DE_adiabatic_solution}
\ee
Since we need to determine $\theta_{\rm{DE}}$, we use the time-time linearised Friedmann equation  
\be
k^{2}\Phi + 3\cH\(\cH\Psi+\Phi'\) = -\frac{3}{2}\cH^{2}\[\Omega_{\rm{r}}\delta_{\rm{r}}+\Omega_{\rm{b}}\delta_{\rm{b}}+\tilde{\Omega}_{\rm{cdm}}\delta_{\rm{cdm}}+\Omega_{\rm{DE}}\delta_{\rm{DE}}\],
\label{Eq:lttf}
\ee
where we only take into consideration radiation (r), baryons (b), CDM and DE. During radiation dominance and under the super-horizon approximation, Eq.~\eqref{Eq:lttf} simplifies to
\be
 \cH\(\cH\Psi+\Phi'\) = -\frac{1}{2}\cH^{2}\delta_{\rm{r}}.
\ee
Moreover, because we are assuming adiabatic perturbations (i.e., $3\delta_{\rm{r}}/4 = \delta_{\rm{cdm}}$) 
\be
 \cH\(\cH\Psi+\Phi'\) = -\frac{2}{3}\cH^{2}\delta_{\rm{cdm}}.
\label{Eq: first_friedmann_equation_adiabatic_solutions}
\ee
We can use likewise the time-space Friedmann equation in radiation dominance
\be
k^{2}\(\cH\Psi+\Phi'\) = 2\cH^{2}\theta_{\rm{r}}.
\label{Eq:tsfe}
\ee
Using Eq.~\eqref{Eq: first_friedmann_equation_adiabatic_solutions} in Eq.~\eqref{Eq:tsfe}, we find  
\be
-\frac{1}{3}\frac{k^{2}}{\cH}\tilde{\delta}_{\rm{cdm}} = \theta_{\rm{r}},
\ee
which in the super-horizon limit (i.e., $k/\mc{H} \rightarrow 0$) yields $\theta_{\rm{r}}\rightarrow 0$.

Then from Eq.~\eqref{eq:DE-velocity-perturbation}, on the super-horizon limit and during radiation dominance (i.e., $r_{\rm{Q}}\rightarrow 0$) the DE velocity perturbation reads
\be
\(1+w_{\rm{DE}}\)\theta_{\rm{DE}} = \frac{k^{2}}{3\cH} \mc{Q}.
\label{Eq: theta_DE_adiabatic_solution}
\ee
By using Eqs.~\eqref{Eq: theta_DE_adiabatic_solution} and ~\eqref{Eq: delta_DE_adiabatic_solution}, we can express the conditions ~\eqref{Eq: adiabatic_initial_condition}-\eqref{Eq: adiabatic_initial_condition_2} in terms of $\mc{Z}$ and $\mc{Q}$ as
\be
\mc{Z} -\mc{Q} = (1+w_{\rm{DE}})\delta_{\rm{cdm}},
\label{Eq: Z_adiabatic_initial_condition}
\ee
\be
\mc{Z}' -\mc{Q}' = (1+w_{\rm{DE}})\delta_{\rm{cdm}}',
\label{Eq:ZQprime}
\ee
where we use that $w_{\rm{DE}}' \rightarrow 0$ during radiation dominance. Eq.~\eqref{Eq:ZQprime} can then be rewritten using Eqs.~\eqref{Eq:Q-prime}, \eqref{Eq:Z-prime}, \eqref{eq:perturbed_energy_for_cdm}, and \eqref{Eq: first_friedmann_equation_adiabatic_solutions}
\be
3\(w_{\rm{DE}}-c_{A}^{2}\)\cH \mc{Z} - 2\cH \mc{Q} = -\(1+w_{\rm{DE}}\)\(\theta_{\rm{cdm}}+2\cH\delta_{\rm{cdm}}+3\cH \Psi\),
\label{Eq: Q_adiabatic_initial_condition}
\ee
taking into account that when radiation dominates $\alpha \rightarrow 4/3$. Finally, the DE adiabatic initial conditions are determined by solving Eqs.~\eqref{Eq: Z_adiabatic_initial_condition} and \eqref{Eq: Q_adiabatic_initial_condition} for $\mc{Z}$ and $\mc{Q}$.

\section{Avoiding ghost and instabilities}
\label{appendix:C}

In Ref.~\cite{Nakamura:2019phn} authors found that in order to avoid scalar ghost, the condition 
\be
a^{2}K_{11} \equiv \frac{3 M_{\rm{pl}}^{2}p^{2}\cH^{2}\Omega_{\rm{DE}}(s+s r_{\rm{Q}} +\tilde{s}\Omega_{\rm{DE}})}{\phi^{2}(-1+p\tilde{s}\Omega_{\rm{DE}})^{2}} >0, ~~~~\tilde{s}=(1-r_{\rm{Q}})s,
\label{Eq: K11}
\ee
must be satisfied. From the vector field equation of motion ~\eqref{Eq: Back vector eom}, the Hubble parameter can be expressed as
\be 
\cH/a \propto (1-r_{\rm{Q}})\phi^{-p}.
\ee
Then we can write the vector field in terms of the DE parameter density $\Omega_{\rm{DE}}$, it reads
\be
\phi \propto (1-r_{\rm{Q}})^{1/p(1+s)}\Omega_{\rm{DE}}^{1/2p(1+s)}.
\ee  
As a result, the first term in $K_{11}$ is proportional to $\left[(1-r_{\rm{Q}})^{2}\Omega_{\rm{DE}}\right]^{(ps-1)/[p(1+s)]}$ and in order to avoid $K_{11}\rightarrow 0$ in the asymptotic past (i.e., when $\Omega_{\rm{DE}}\rightarrow 0$), the parameters $p$ and $s$ must satisfy 
\be
\frac{ps-1}{p(1+s)} \leq 0.
\label{Eq: constrain 1}
\ee
During radiation dominance we assume $r_{\rm{Q}},\, \Omega_{\rm{DE}} \rightarrow 0$, so that the condition in Eq.~\eqref{Eq: K11} restricts $s$ to positive values (i.e., $s>0$). The constraint~\eqref{Eq: constrain 1} along with $s>0$ implies $p>0$ and
\be
0 < p_2 \leq 1.
\label{Eq:p2}
\ee
Following Ref.~\cite{Nakamura:2019phn}, in order to avoid Laplacian instability in the scalar perturbations, the squared propagation speed of vector field perturbations $c_{S}^{2}$ must be positive, namely, 
\begin{align}
c_{S}^{2} &= \frac{\tilde{s}\(p(5+6s)-3\)\(2s-\tilde{s}\)+s p^{2}\tilde{s}^{4}\Omega_{\rm{DE}}^{2} +\tilde{s}^{2}\left[3-p\(3+(6+4p)s\) + 2p\( 1+p\)\tilde{s}\right]\Omega_{\rm{DE}}}{6p^{2}\(2s-\tilde{s}+\tilde{s}^{2}\Omega_{\rm{DE}}\)^{2}} \nonumber \\
&+ \frac{\tilde{s}^{2}\(2ps+p-1\)\Omega_{\rm{r}}}{6p^{2}\(2s-\tilde{s}+\tilde{s}^{2}\Omega_{\rm{DE}}\)^{2}} + \frac{2M_{\rm{pl}}^{2}\tilde{s}^{2}\Omega_{\rm{DE}}}{3G_{2,F}\phi^{2}\(2s-\tilde{s}+\tilde{s}^{2}\Omega_{\rm{DE}}\)} > 0.
\label{Eq:cs2}
\end{align}
By analysing the last term in Eq.~\eqref{Eq:cs2}, which is proportional to $\Omega_{\rm{DE}}^{[p(1+s)-1]/[p(1+s)]}$, it becomes clear that as long as $[p(1+s)-1]/[p(1+s)] > 0$ the constraint~\eqref{Eq:cs2} is satisfied when $\Omega_{\rm{DE}}\rightarrow 0$. Since $s > 0$ and $p > 0$, it is required that 
\begin{equation}
p_2(1+s) > s.
\end{equation}
During radiation, matter, and DE dominance epochs, $c_{S}^{2}$ is respectively given by
\begin{equation}
c_{S}^{2~(\rm{r})} = \frac{-2+p(3+4s)}{3p^{2}}, \quad c_{S}^{2 (\rm{m})} = \frac{-3+p(5+6s)}{6p^{2}}\frac{1-r_{\rm{Q}}^{(\rm{m})}}{1+r_{\rm{Q}}^{(\rm{m})}}, \quad c_{S}^{2(\rm{DE})} = \frac{1-ps}{3p(1+s)}.
\end{equation}
Note that $-3+p(5+6s)>0$ when the parameter $s$ satisfies
\be
0<s<-\frac{p_2}{-1+p_2}.
\label{Eq:s}
\ee 
Under the constraint~\eqref{Eq:s}, $c_{S}^{2~(\rm{r})}>0$ and the condition $c_{S}^{2~(\rm{m})}>0$ requires $-1<r_{\rm{Q}}^{(\rm{m})}<1$.

We now focus on the equation of motion governing $r_{\mathrm{Q}}$
\be
r_{\mathrm{Q}}' = \cH r_{\mathrm{Q}} \left[ \frac{q - p_{2}}{p_{2}} \epsilon_{\phi} - 3 \right],
\ee
which we obtain by differentiating Eq.~\eqref{Eq: G3X in Q} while taking into account Eq.~\eqref{Eq: eps_phi}. Note that during the matter-dominated era (i.e., $\Omega_{\rm{r}},\, \Omega_{\rm{DE}} \rightarrow 0$) 
\be
r'^{\rm{(m)}}_{\rm{Q}} =  \cH^{\rm{(m)}} r_{\rm{Q}}^{\rm{(m)}} \left[\frac{q-p_{2}}{p_{2}}3s-3\right].
\ee
If we take into consideration Eq.~\eqref{Eq:param-def}, we realise
\be 
q=2p_{3}-p_{2}+1,
\label{Eq:q-constraint}
\ee
implies $r'^{\rm{(m)}}_{\rm{Q}} = 0$. By exploring the equation of motion \eqref{Eq: Back vector eom} during matter dominance, we can determine the upper limit $ r_{\rm{Q}}^{\mathrm{(m)}} < 1$. Reorganizing terms in Eq.~\eqref{Eq: Back vector eom}, we obtain
\begin{equation}
1 + \frac{3}{2^{p_{3}-p_{2}}b_{2}} \frac{\mathcal{H}^{\mathrm{(m)}}}{a} \phi^{\mathrm{(m)}} \left(\phi^{\mathrm{(m)}}\right)^{2(p_{3}-p_{2})} - r_{\rm{Q}}^{\mathrm{(m)}} = 0.
\label{Eq: G3X in Q in m}
\end{equation}   
Using Eq.~\eqref{Eq: eps_phi}, we find the vector field behaves as 
\be 
\phi^{\mathrm{(m)}} = \phi_{0}^{\mathrm{(m)}} a^{\frac{3}{2}\frac{s}{p_{2}}},
\label{Eq:pmd}
\ee
where $\phi_{0}^{\mathrm{(m)}}$ is an integration constant. Additionally, the conformal Hubble parameter behaves as in Eq.~\eqref{Eq:HMD}. Substituting Eqs.~\eqref{Eq:pmd} and \eqref{Eq:HMD} into \eqref{Eq: G3X in Q in m} and regarding $b_{2}<0$, we get
\begin{equation}
r_{\rm{Q}}^{\mathrm{(m)}} = 1 - \frac{3}{2^{p_{3}-p_{2}}\lvert b_{2}\rvert} H_0 \sqrt{\Omega_{\mathrm{cdm}}^{0}}(\phi_{0}^{\mathrm{(m)}})^p < 1. 
\end{equation} 

\acknowledgments

This work was supported by Patrimonio Aut\'onomo - Fondo Nacional de Financiamiento para la Ciencia, la Tecnolog\'ia y la Innovaci\'on Francisco Jos\'e de Caldas (MINCIENCIAS - COLOMBIA) Grant No. 110685269447 RC-80740-465-2020, projects 69475 and 69723.

\section*{Numerical codes}

Modified \texttt{CLASS} code reproducing results in this work can be found in the GitHub branch \texttt{cvde} of the repository  \href{https://github.com/wilmarcardonac/EFCLASS.git}{\texttt{EFCLASS}}. Modified \texttt{MontePython} code can be found in the repository \href{https://github.com/wilmarcardonac/montepython_public.git}{\texttt{montepython\_public}}.

\bibliographystyle{JHEPmodplain}
\bibliography{paper}

\providecommand{\href}[2]{#2}\begingroup\raggedright\begin{thebibliography}{10}

\bibitem{Planck:2018vyg}
{\bf Planck} Collaboration, N.~Aghanim {\em et~al.}, {\it {Planck 2018 results.
  VI. Cosmological parameters}},  {\sl Astron. Astrophys.} {\bf 641} (2020) A6,
  [\href{http://arxiv.org/abs/1807.06209}{{\sf arXiv:1807.06209}}],
  [\href{http://dx.doi.org/10.1051/0004-6361/201833910}{{\sf
  doi:10.1051/0004-6361/201833910}}]. [Erratum: Astron.Astrophys. 652, C4
  (2021)].

\bibitem{PhysRevLett.122.171301}
{\bf DES Collaboration} Collaboration, T.~M.~C. Abbott {\em et~al.}, {\it
  Cosmological constraints from multiple probes in the dark energy survey},
  {\sl Phys. Rev. Lett.} {\bf 122} (May, 2019) 171301,
  [\href{http://dx.doi.org/10.1103/PhysRevLett.122.171301}{{\sf
  doi:10.1103/PhysRevLett.122.171301}}].

\bibitem{Sawala:2022xom}
T.~Sawala, M.~Cautun, C.~S. Frenk, J.~Helly, J.~Jasche, A.~Jenkins, P.~H.
  Johansson, G.~Lavaux, S.~McAlpine, and M.~Schaller, {\it {The Milky
  Way\textquoteright{}s plane of satellites is consistent with
  \ensuremath{\Lambda}CDM}},  {\sl Nature Astron.} {\bf 7} (2023), no.~4
  481--491, [\href{http://arxiv.org/abs/2205.02860}{{\sf arXiv:2205.02860}}],
  [\href{http://dx.doi.org/10.1038/s41550-022-01856-z}{{\sf
  doi:10.1038/s41550-022-01856-z}}].

\bibitem{Heavens:2017hkr}
A.~Heavens, Y.~Fantaye, E.~Sellentin, H.~Eggers, Z.~Hosenie, S.~Kroon, and
  A.~Mootoovaloo, {\it {No evidence for extensions to the standard cosmological
  model}},  {\sl Phys. Rev. Lett.} {\bf 119} (2017), no.~10 101301,
  [\href{http://arxiv.org/abs/1704.03467}{{\sf arXiv:1704.03467}}],
  [\href{http://dx.doi.org/10.1103/PhysRevLett.119.101301}{{\sf
  doi:10.1103/PhysRevLett.119.101301}}].

\bibitem{Dutta:2019pio}
K.~Dutta, A.~Roy, Ruchika, A.~A. Sen, and M.~M. Sheikh-Jabbari, {\it {Cosmology
  with low-redshift observations: No signal for new physics}},  {\sl Phys. Rev.
  D} {\bf 100} (2019), no.~10 103501,
  [\href{http://arxiv.org/abs/1908.07267}{{\sf arXiv:1908.07267}}],
  [\href{http://dx.doi.org/10.1103/PhysRevD.100.103501}{{\sf
  doi:10.1103/PhysRevD.100.103501}}].

\bibitem{Weinberg:1988cp}
S.~Weinberg, {\it {The Cosmological Constant Problem}},  {\sl Rev. Mod. Phys.}
  {\bf 61} (1989) 1--23, [\href{http://dx.doi.org/10.1103/RevModPhys.61.1}{{\sf
  doi:10.1103/RevModPhys.61.1}}]. [,569(1988)].

\bibitem{Carroll:2000fy}
S.~M. Carroll, {\it {The Cosmological constant}},  {\sl Living Rev. Rel.} {\bf
  4} (2001) 1, [\href{http://arxiv.org/abs/astro-ph/0004075}{{\sf
  arXiv:astro-ph/0004075}}], [\href{http://dx.doi.org/10.12942/lrr-2001-1}{{\sf
  doi:10.12942/lrr-2001-1}}].

\bibitem{Bertone:2016nfn}
G.~Bertone and D.~Hooper, {\it {History of dark matter}},  {\sl Rev. Mod.
  Phys.} {\bf 90} (2018), no.~4 045002,
  [\href{http://arxiv.org/abs/1605.04909}{{\sf arXiv:1605.04909}}],
  [\href{http://dx.doi.org/10.1103/RevModPhys.90.045002}{{\sf
  doi:10.1103/RevModPhys.90.045002}}].

\bibitem{Martin:2012bt}
J.~Martin, {\it {Everything You Always Wanted To Know About The Cosmological
  Constant Problem (But Were Afraid To Ask)}},  {\sl Comptes Rendus Physique}
  {\bf 13} (2012) 566--665, [\href{http://arxiv.org/abs/1205.3365}{{\sf
  arXiv:1205.3365}}], [\href{http://dx.doi.org/10.1016/j.crhy.2012.04.008}{{\sf
  doi:10.1016/j.crhy.2012.04.008}}].

\bibitem{PhysRevLett.127.161302}
C.~Skordis and T.~Z\l{}o\ifmmode~\acute{s}\else \'{s}\fi{}nik, {\it New
  relativistic theory for modified newtonian dynamics},  {\sl Phys. Rev. Lett.}
  {\bf 127} (Oct, 2021) 161302,
  [\href{http://dx.doi.org/10.1103/PhysRevLett.127.161302}{{\sf
  doi:10.1103/PhysRevLett.127.161302}}].

\bibitem{Bertone:2004pz}
G.~Bertone, D.~Hooper, and J.~Silk, {\it {Particle dark matter: Evidence,
  candidates and constraints}},  {\sl Phys. Rept.} {\bf 405} (2005) 279--390,
  [\href{http://arxiv.org/abs/hep-ph/0404175}{{\sf arXiv:hep-ph/0404175}}],
  [\href{http://dx.doi.org/10.1016/j.physrep.2004.08.031}{{\sf
  doi:10.1016/j.physrep.2004.08.031}}].

\bibitem{Copeland:2006wr}
E.~J. Copeland, M.~Sami, and S.~Tsujikawa, {\it {Dynamics of dark energy}},
  {\sl Int. J. Mod. Phys.} {\bf D15} (2006) 1753--1936,
  [\href{http://arxiv.org/abs/hep-th/0603057}{{\sf arXiv:hep-th/0603057}}],
  [\href{http://dx.doi.org/10.1142/S021827180600942X}{{\sf
  doi:10.1142/S021827180600942X}}].

\bibitem{Clifton:2011jh}
T.~Clifton, P.~G. Ferreira, A.~Padilla, and C.~Skordis, {\it {Modified Gravity
  and Cosmology}},  {\sl Phys. Rept.} {\bf 513} (2012) 1--189,
  [\href{http://arxiv.org/abs/1106.2476}{{\sf arXiv:1106.2476}}],
  [\href{http://dx.doi.org/10.1016/j.physrep.2012.01.001}{{\sf
  doi:10.1016/j.physrep.2012.01.001}}].

\bibitem{Bamba:2012cp}
K.~Bamba, S.~Capozziello, S.~Nojiri, and S.~D. Odintsov, {\it {Dark energy
  cosmology: the equivalent description via different theoretical models and
  cosmography tests}},  {\sl Astrophys. Space Sci.} {\bf 342} (2012) 155--228,
  [\href{http://arxiv.org/abs/1205.3421}{{\sf arXiv:1205.3421}}],
  [\href{http://dx.doi.org/10.1007/s10509-012-1181-8}{{\sf
  doi:10.1007/s10509-012-1181-8}}].

\bibitem{Riess:2019cxk}
A.~G. Riess, S.~Casertano, W.~Yuan, L.~M. Macri, and D.~Scolnic, {\it {Large
  Magellanic Cloud} {Cepheid Standards Provide a 1\% Foundation for the
  Determination of the Hubble Constant} and {Stronger Evidence for Physics}
  beyond {$\Lambda$CDM}},  {\sl Astrophys. J.} {\bf 876} (2019), no.~1 85,
  [\href{http://arxiv.org/abs/1903.07603}{{\sf arXiv:1903.07603}}],
  [\href{http://dx.doi.org/10.3847/1538-4357/ab1422}{{\sf
  doi:10.3847/1538-4357/ab1422}}].

\bibitem{Freedman:2021ahq}
W.~L. Freedman, {\it {Measurements of the Hubble Constant: Tensions in
  Perspective}},  {\sl Astrophys. J.} {\bf 919} (2021), no.~1 16,
  [\href{http://arxiv.org/abs/2106.15656}{{\sf arXiv:2106.15656}}],
  [\href{http://dx.doi.org/10.3847/1538-4357/ac0e95}{{\sf
  doi:10.3847/1538-4357/ac0e95}}].

\bibitem{Riess:2021jrx}
A.~G. Riess {\em et~al.}, {\it {A Comprehensive Measurement of the Local Value
  of the Hubble Constant with 1 km s$^{−1}$ Mpc$^{−1}$ Uncertainty from the
  Hubble Space Telescope and the SH0ES Team}},  {\sl Astrophys. J. Lett.} {\bf
  934} (2022), no.~1 L7, [\href{http://arxiv.org/abs/2112.04510}{{\sf
  arXiv:2112.04510}}], [\href{http://dx.doi.org/10.3847/2041-8213/ac5c5b}{{\sf
  doi:10.3847/2041-8213/ac5c5b}}].

\bibitem{10.1093/mnras/stac2429}
A.~Amon and G.~Efstathiou, {\it {A non-linear solution to the S8 tension?}},
  {\sl Monthly Notices of the Royal Astronomical Society} (09, 2022)
  [\href{http://arxiv.org/abs/https://academic.oup.com/mnras/advance-article-pdf/doi/10.1093/mnras/stac2429/45688270/stac2429.pdf}{{\sf
  https://academic.oup.com/mnras/advance-article-pdf/doi/10.1093/mnras/stac2429/45688270/stac2429.pdf}}],
  [\href{http://dx.doi.org/10.1093/mnras/stac2429}{{\sf
  doi:10.1093/mnras/stac2429}}]. stac2429.

\bibitem{Asghari:2019qld}
M.~Asghari, J.~Beltr\'an~Jim\'enez, S.~Khosravi, and D.~F. Mota, {\it {On
  structure formation from a small-scales-interacting dark sector}},  {\sl
  JCAP} {\bf 04} (2019) 042, [\href{http://arxiv.org/abs/1902.05532}{{\sf
  arXiv:1902.05532}}],
  [\href{http://dx.doi.org/10.1088/1475-7516/2019/04/042}{{\sf
  doi:10.1088/1475-7516/2019/04/042}}].

\bibitem{Cardona:2022mdq}
W.~Cardona and D.~Figueruelo, {\it {Momentum transfer in the dark sector and
  lensing convergence in upcoming galaxy surveys}},  {\sl JCAP} {\bf 12} (2022)
  010, [\href{http://arxiv.org/abs/2209.12583}{{\sf arXiv:2209.12583}}],
  [\href{http://dx.doi.org/10.1088/1475-7516/2022/12/010}{{\sf
  doi:10.1088/1475-7516/2022/12/010}}].

\bibitem{PhysRevLett.131.111001}
N.-M. Nguyen, D.~Huterer, and Y.~Wen, {\it Evidence for suppression of
  structure growth in the concordance cosmological model},  {\sl Phys. Rev.
  Lett.} {\bf 131} (Sep, 2023) 111001,
  [\href{http://dx.doi.org/10.1103/PhysRevLett.131.111001}{{\sf
  doi:10.1103/PhysRevLett.131.111001}}].

\bibitem{Efstathiou:2013via}
G.~Efstathiou, {\it {H0 Revisited}},  {\sl Mon. Not. Roy. Astron. Soc.} {\bf
  440} (2014), no.~2 1138--1152, [\href{http://arxiv.org/abs/1311.3461}{{\sf
  arXiv:1311.3461}}], [\href{http://dx.doi.org/10.1093/mnras/stu278}{{\sf
  doi:10.1093/mnras/stu278}}].

\bibitem{Mazo:2022auo}
B.~Y. D.~V. Mazo, A.~E. Romano, and M.~A.~C. Quintero, {\it {$H_0$ tension or
  $M$ overestimation?}},  {\sl Eur. Phys. J. C} {\bf 82} (2022), no.~7 610,
  [\href{http://arxiv.org/abs/2202.11852}{{\sf arXiv:2202.11852}}],
  [\href{http://dx.doi.org/10.1140/epjc/s10052-022-10526-3}{{\sf
  doi:10.1140/epjc/s10052-022-10526-3}}].

\bibitem{Dainotti:2021pqg}
M.~G. Dainotti, B.~De~Simone, T.~Schiavone, G.~Montani, E.~Rinaldi, and
  G.~Lambiase, {\it {On the Hubble constant tension in the SNe Ia Pantheon
  sample}},  {\sl Astrophys. J.} {\bf 912} (2021), no.~2 150,
  [\href{http://arxiv.org/abs/2103.02117}{{\sf arXiv:2103.02117}}],
  [\href{http://dx.doi.org/10.3847/1538-4357/abeb73}{{\sf
  doi:10.3847/1538-4357/abeb73}}].

\bibitem{Dainotti:2022bzg}
M.~G. Dainotti, B.~De~Simone, T.~Schiavone, G.~Montani, E.~Rinaldi,
  G.~Lambiase, M.~Bogdan, and S.~Ugale, {\it {On the Evolution of the Hubble
  Constant with the SNe Ia Pantheon Sample and Baryon Acoustic Oscillations: A
  Feasibility Study for GRB-Cosmology in 2030}},  {\sl Galaxies} {\bf 10}
  (2022), no.~1 24, [\href{http://arxiv.org/abs/2201.09848}{{\sf
  arXiv:2201.09848}}], [\href{http://dx.doi.org/10.3390/galaxies10010024}{{\sf
  doi:10.3390/galaxies10010024}}].

\bibitem{Arjona:2018jhh}
R.~Arjona, W.~Cardona, and S.~Nesseris, {\it {Unraveling the effective fluid
  approach for $f(R)$ models in the subhorizon approximation}},  {\sl Phys.
  Rev.} {\bf D99} (2019), no.~4 043516,
  [\href{http://arxiv.org/abs/1811.02469}{{\sf arXiv:1811.02469}}],
  [\href{http://dx.doi.org/10.1103/PhysRevD.99.043516}{{\sf
  doi:10.1103/PhysRevD.99.043516}}].

\bibitem{Arjona:2019rfn}
R.~Arjona, W.~Cardona, and S.~Nesseris, {\it {Designing Horndeski and the
  effective fluid approach}},  {\sl Phys. Rev. D} {\bf 100} (2019), no.~6
  063526, [\href{http://arxiv.org/abs/1904.06294}{{\sf arXiv:1904.06294}}],
  [\href{http://dx.doi.org/10.1103/PhysRevD.100.063526}{{\sf
  doi:10.1103/PhysRevD.100.063526}}].

\bibitem{Cardona:2022lcz}
W.~Cardona, J.~B. Orjuela-Quintana, and C.~A. Valenzuela-Toledo, {\it {An
  effective fluid description of scalar-vector-tensor theories under the
  sub-horizon and quasi-static approximations}},  {\sl JCAP} {\bf 08} (2022),
  no.~08 059, [\href{http://arxiv.org/abs/2206.02895}{{\sf arXiv:2206.02895}}],
  [\href{http://dx.doi.org/10.1088/1475-7516/2022/08/059}{{\sf
  doi:10.1088/1475-7516/2022/08/059}}].

\bibitem{Geng:2021jso}
C.-Q. Geng, Y.-T. Hsu, J.-R. Lu, and L.~Yin, {\it {A Dark Energy model from
  Generalized Proca Theory}},  {\sl Phys. Dark Univ.} {\bf 32} (2021) 100819,
  [\href{http://arxiv.org/abs/2104.06577}{{\sf arXiv:2104.06577}}],
  [\href{http://dx.doi.org/10.1016/j.dark.2021.100819}{{\sf
  doi:10.1016/j.dark.2021.100819}}].

\bibitem{Nakamura:2018oyy}
S.~Nakamura, A.~De~Felice, R.~Kase, and S.~Tsujikawa, {\it {Constraints on
  massive vector dark energy models from integrated Sachs-Wolfe-galaxy
  cross-correlations}},  {\sl Phys. Rev. D} {\bf 99} (2019), no.~6 063533,
  [\href{http://arxiv.org/abs/1811.07541}{{\sf arXiv:1811.07541}}],
  [\href{http://dx.doi.org/10.1103/PhysRevD.99.063533}{{\sf
  doi:10.1103/PhysRevD.99.063533}}].

\bibitem{deFelice:2017paw}
A.~de~Felice, L.~Heisenberg, and S.~Tsujikawa, {\it {Observational constraints
  on generalized Proca theories}},  {\sl Phys. Rev. D} {\bf 95} (2017), no.~12
  123540, [\href{http://arxiv.org/abs/1703.09573}{{\sf arXiv:1703.09573}}],
  [\href{http://dx.doi.org/10.1103/PhysRevD.95.123540}{{\sf
  doi:10.1103/PhysRevD.95.123540}}].

\bibitem{DeFelice:2016yws}
A.~De~Felice, L.~Heisenberg, R.~Kase, S.~Mukohyama, S.~Tsujikawa, and Y.-l.
  Zhang, {\it {Cosmology in generalized Proca theories}},  {\sl JCAP} {\bf 06}
  (2016) 048, [\href{http://arxiv.org/abs/1603.05806}{{\sf arXiv:1603.05806}}],
  [\href{http://dx.doi.org/10.1088/1475-7516/2016/06/048}{{\sf
  doi:10.1088/1475-7516/2016/06/048}}].

\bibitem{Armendariz-Picon:2004say}
C.~Armendariz-Picon, {\it {Could dark energy be vector-like?}},  {\sl JCAP}
  {\bf 07} (2004) 007, [\href{http://arxiv.org/abs/astro-ph/0405267}{{\sf
  arXiv:astro-ph/0405267}}],
  [\href{http://dx.doi.org/10.1088/1475-7516/2004/07/007}{{\sf
  doi:10.1088/1475-7516/2004/07/007}}].

\bibitem{Koivisto:2008xf}
T.~Koivisto and D.~F. Mota, {\it {Vector Field Models of Inflation and Dark
  Energy}},  {\sl JCAP} {\bf 08} (2008) 021,
  [\href{http://arxiv.org/abs/0805.4229}{{\sf arXiv:0805.4229}}],
  [\href{http://dx.doi.org/10.1088/1475-7516/2008/08/021}{{\sf
  doi:10.1088/1475-7516/2008/08/021}}].

\bibitem{Gomez:2020sfz}
L.~G. Gomez and Y.~Rodriguez, {\it {Coupled multi-Proca vector dark energy}},
  {\sl Phys. Dark Univ.} {\bf 31} (2021) 100759,
  [\href{http://arxiv.org/abs/2004.06466}{{\sf arXiv:2004.06466}}],
  [\href{http://dx.doi.org/10.1016/j.dark.2020.100759}{{\sf
  doi:10.1016/j.dark.2020.100759}}].

\bibitem{Alvarez:2019ues}
M.~\'Alvarez, J.~B. Orjuela-Quintana, Y.~Rodriguez, and C.~A.
  Valenzuela-Toledo, {\it {Einstein Yang\textendash{}Mills Higgs dark energy
  revisited}},  {\sl Class. Quant. Grav.} {\bf 36} (2019), no.~19 195004,
  [\href{http://arxiv.org/abs/1901.04624}{{\sf arXiv:1901.04624}}],
  [\href{http://dx.doi.org/10.1088/1361-6382/ab3775}{{\sf
  doi:10.1088/1361-6382/ab3775}}].

\bibitem{Zuntz:2010jp}
J.~Zuntz, T.~G. Zlosnik, F.~Bourliot, P.~G. Ferreira, and G.~D. Starkman, {\it
  {Vector field models of modified gravity and the dark sector}},  {\sl Phys.
  Rev. D} {\bf 81} (2010) 104015, [\href{http://arxiv.org/abs/1002.0849}{{\sf
  arXiv:1002.0849}}], [\href{http://dx.doi.org/10.1103/PhysRevD.81.104015}{{\sf
  doi:10.1103/PhysRevD.81.104015}}].

\bibitem{BeltranJimenez:2008iye}
J.~Beltran~Jimenez and A.~L. Maroto, {\it {A cosmic vector for dark energy}},
  {\sl Phys. Rev. D} {\bf 78} (2008) 063005,
  [\href{http://arxiv.org/abs/0801.1486}{{\sf arXiv:0801.1486}}],
  [\href{http://dx.doi.org/10.1103/PhysRevD.78.063005}{{\sf
  doi:10.1103/PhysRevD.78.063005}}].

\bibitem{Bamba:2008ja}
K.~Bamba and S.~D. Odintsov, {\it {Inflation and late-time cosmic acceleration
  in non-minimal Maxwell-$F(R)$ gravity and the generation of large-scale
  magnetic fields}},  {\sl JCAP} {\bf 04} (2008) 024,
  [\href{http://arxiv.org/abs/0801.0954}{{\sf arXiv:0801.0954}}],
  [\href{http://dx.doi.org/10.1088/1475-7516/2008/04/024}{{\sf
  doi:10.1088/1475-7516/2008/04/024}}].

\bibitem{Bamba:2008xa}
K.~Bamba, S.~Nojiri, and S.~D. Odintsov, {\it {Inflationary cosmology and the
  late-time accelerated expansion of the universe in non-minimal
  Yang-Mills-F(R) gravity and non-minimal vector-F(R) gravity}},  {\sl Phys.
  Rev. D} {\bf 77} (2008) 123532, [\href{http://arxiv.org/abs/0803.3384}{{\sf
  arXiv:0803.3384}}], [\href{http://dx.doi.org/10.1103/PhysRevD.77.123532}{{\sf
  doi:10.1103/PhysRevD.77.123532}}].

\bibitem{Heisenberg:2020xak}
L.~Heisenberg and H.~Villarrubia-Rojo, {\it {Proca in the sky}},  {\sl JCAP}
  {\bf 03} (2021) 032, [\href{http://arxiv.org/abs/2010.00513}{{\sf
  arXiv:2010.00513}}],
  [\href{http://dx.doi.org/10.1088/1475-7516/2021/03/032}{{\sf
  doi:10.1088/1475-7516/2021/03/032}}].

\bibitem{Amendola:1999er}
L.~Amendola, {\it {Coupled quintessence}},  {\sl Phys. Rev. D} {\bf 62} (2000)
  043511, [\href{http://arxiv.org/abs/astro-ph/9908023}{{\sf
  arXiv:astro-ph/9908023}}],
  [\href{http://dx.doi.org/10.1103/PhysRevD.62.043511}{{\sf
  doi:10.1103/PhysRevD.62.043511}}].

\bibitem{Wang:2016lxa}
B.~Wang, E.~Abdalla, F.~Atrio-Barandela, and D.~Pavon, {\it {Dark Matter and
  Dark Energy Interactions: Theoretical Challenges, Cosmological Implications
  and Observational Signatures}},  {\sl Rept. Prog. Phys.} {\bf 79} (2016),
  no.~9 096901, [\href{http://arxiv.org/abs/1603.08299}{{\sf
  arXiv:1603.08299}}],
  [\href{http://dx.doi.org/10.1088/0034-4885/79/9/096901}{{\sf
  doi:10.1088/0034-4885/79/9/096901}}].

\bibitem{Nakamura:2019phn}
S.~Nakamura, R.~Kase, and S.~Tsujikawa, {\it {Coupled vector dark energy}},
  {\sl JCAP} {\bf 12} (2019) 032, [\href{http://arxiv.org/abs/1907.12216}{{\sf
  arXiv:1907.12216}}],
  [\href{http://dx.doi.org/10.1088/1475-7516/2019/12/032}{{\sf
  doi:10.1088/1475-7516/2019/12/032}}].

\bibitem{Damour:1994zq}
T.~Damour and A.~M. Polyakov, {\it {The String dilaton and a least coupling
  principle}},  {\sl Nucl. Phys. B} {\bf 423} (1994) 532--558,
  [\href{http://arxiv.org/abs/hep-th/9401069}{{\sf arXiv:hep-th/9401069}}],
  [\href{http://dx.doi.org/10.1016/0550-3213(94)90143-0}{{\sf
  doi:10.1016/0550-3213(94)90143-0}}].

\bibitem{Hogg:2004vw}
D.~W. Hogg, D.~J. Eisenstein, M.~R. Blanton, N.~A. Bahcall, J.~Brinkmann, J.~E.
  Gunn, and D.~P. Schneider, {\it {Cosmic homogeneity demonstrated with
  luminous red galaxies}},  {\sl Astrophys. J.} {\bf 624} (2005) 54--58,
  [\href{http://arxiv.org/abs/astro-ph/0411197}{{\sf arXiv:astro-ph/0411197}}],
  [\href{http://dx.doi.org/10.1086/429084}{{\sf doi:10.1086/429084}}].

\bibitem{Ade:2015hxq}
{\bf Planck} Collaboration, P.~A.~R. Ade {\em et~al.}, {\it {Planck 2015
  results. XVI. Isotropy and statistics of the CMB}},  {\sl Astron. Astrophys.}
  {\bf 594} (2016) A16, [\href{http://arxiv.org/abs/1506.07135}{{\sf
  arXiv:1506.07135}}],
  [\href{http://dx.doi.org/10.1051/0004-6361/201526681}{{\sf
  doi:10.1051/0004-6361/201526681}}].

\bibitem{Marinoni:2012ba}
C.~Marinoni, J.~Bel, and A.~Buzzi, {\it {The Scale of Cosmic Isotropy}},  {\sl
  JCAP} {\bf 10} (2012) 036, [\href{http://arxiv.org/abs/1205.3309}{{\sf
  arXiv:1205.3309}}],
  [\href{http://dx.doi.org/10.1088/1475-7516/2012/10/036}{{\sf
  doi:10.1088/1475-7516/2012/10/036}}].

\bibitem{PhysRevLett.117.131302}
D.~Saadeh, S.~M. Feeney, A.~Pontzen, H.~V. Peiris, and J.~D. McEwen, {\it How
  isotropic is the universe?},  {\sl Phys. Rev. Lett.} {\bf 117} (Sep, 2016)
  131302, [\href{http://dx.doi.org/10.1103/PhysRevLett.117.131302}{{\sf
  doi:10.1103/PhysRevLett.117.131302}}].

\bibitem{Koivisto:2005nr}
T.~Koivisto, {\it {Growth of perturbations in dark matter coupled with
  quintessence}},  {\sl Phys. Rev. D} {\bf 72} (2005) 043516,
  [\href{http://arxiv.org/abs/astro-ph/0504571}{{\sf arXiv:astro-ph/0504571}}],
  [\href{http://dx.doi.org/10.1103/PhysRevD.72.043516}{{\sf
  doi:10.1103/PhysRevD.72.043516}}].

\bibitem{Avelino:2018qgt}
P.~P. Avelino and L.~Sousa, {\it {Matter Lagrangian of particles and fluids}},
  {\sl Phys. Rev. D} {\bf 97} (2018), no.~6 064019,
  [\href{http://arxiv.org/abs/1802.03961}{{\sf arXiv:1802.03961}}],
  [\href{http://dx.doi.org/10.1103/PhysRevD.97.064019}{{\sf
  doi:10.1103/PhysRevD.97.064019}}].

\bibitem{Blas:2011rf}
D.~Blas, J.~Lesgourgues, and T.~Tram, {\it {The Cosmic Linear Anisotropy
  Solving System (CLASS) II: Approximation schemes}},  {\sl JCAP} {\bf 1107}
  (2011) 034, [\href{http://arxiv.org/abs/1104.2933}{{\sf arXiv:1104.2933}}],
  [\href{http://dx.doi.org/10.1088/1475-7516/2011/07/034}{{\sf
  doi:10.1088/1475-7516/2011/07/034}}].

\bibitem{Sapone:2009mb}
D.~Sapone and M.~Kunz, {\it {Fingerprinting Dark Energy}},  {\sl Phys. Rev.}
  {\bf D80} (2009) 083519, [\href{http://arxiv.org/abs/0909.0007}{{\sf
  arXiv:0909.0007}}], [\href{http://dx.doi.org/10.1103/PhysRevD.80.083519}{{\sf
  doi:10.1103/PhysRevD.80.083519}}].

\bibitem{Pan-STARRS1:2017jku}
{\bf Pan-STARRS1} Collaboration, D.~M. Scolnic {\em et~al.}, {\it {The Complete
  Light-curve Sample of Spectroscopically Confirmed SNe Ia from Pan-STARRS1 and
  Cosmological Constraints from the Combined Pantheon Sample}},  {\sl
  Astrophys. J.} {\bf 859} (2018), no.~2 101,
  [\href{http://arxiv.org/abs/1710.00845}{{\sf arXiv:1710.00845}}],
  [\href{http://dx.doi.org/10.3847/1538-4357/aab9bb}{{\sf
  doi:10.3847/1538-4357/aab9bb}}].

\bibitem{BOSS:2016wmc}
{\bf BOSS} Collaboration, S.~Alam {\em et~al.}, {\it {The clustering of
  galaxies in the completed SDSS-III Baryon Oscillation Spectroscopic Survey:
  cosmological analysis of the DR12 galaxy sample}},  {\sl Mon. Not. Roy.
  Astron. Soc.} {\bf 470} (2017), no.~3 2617--2652,
  [\href{http://arxiv.org/abs/1607.03155}{{\sf arXiv:1607.03155}}],
  [\href{http://dx.doi.org/10.1093/mnras/stx721}{{\sf
  doi:10.1093/mnras/stx721}}].

\bibitem{2011}
F.~Beutler, C.~Blake, M.~Colless, D.~H. Jones, L.~Staveley-Smith, L.~Campbell,
  Q.~Parker, W.~Saunders, and F.~Watson, {\it The 6df galaxy survey: baryon
  acoustic oscillations and the local hubble constant},  {\sl Monthly Notices
  of the Royal Astronomical Society} {\bf 416} (Jul, 2011) 3017–3032,
  [\href{http://dx.doi.org/10.1111/j.1365-2966.2011.19250.x}{{\sf
  doi:10.1111/j.1365-2966.2011.19250.x}}].

\bibitem{Ross:2014qpa}
A.~J. Ross, L.~Samushia, C.~Howlett, W.~J. Percival, A.~Burden, and M.~Manera,
  {\it {The clustering of the SDSS DR7 main Galaxy sample \textendash{} I. A 4
  per cent distance measure at $z = 0.15$}},  {\sl Mon. Not. Roy. Astron. Soc.}
  {\bf 449} (2015), no.~1 835--847, [\href{http://arxiv.org/abs/1409.3242}{{\sf
  arXiv:1409.3242}}], [\href{http://dx.doi.org/10.1093/mnras/stv154}{{\sf
  doi:10.1093/mnras/stv154}}].

\bibitem{Audren:2012wb}
B.~Audren, J.~Lesgourgues, K.~Benabed, and S.~Prunet, {\it {Conservative
  Constraints on Early Cosmology: an illustration of the Monte Python
  cosmological parameter inference code}},  {\sl JCAP} {\bf 1302} (2013) 001,
  [\href{http://arxiv.org/abs/1210.7183}{{\sf arXiv:1210.7183}}],
  [\href{http://dx.doi.org/10.1088/1475-7516/2013/02/001}{{\sf
  doi:10.1088/1475-7516/2013/02/001}}].

\bibitem{Brinckmann:2018cvx}
T.~Brinckmann and J.~Lesgourgues, {\it {MontePython 3: boosted MCMC sampler and
  other features}},  {\sl Phys. Dark Univ.} {\bf 24} (2019) 100260,
  [\href{http://arxiv.org/abs/1804.07261}{{\sf arXiv:1804.07261}}],
  [\href{http://dx.doi.org/10.1016/j.dark.2018.100260}{{\sf
  doi:10.1016/j.dark.2018.100260}}].

\bibitem{Lewis:2002ah}
A.~Lewis and S.~Bridle, {\it {Cosmological parameters from CMB and other data:
  A Monte Carlo approach}},  {\sl Phys. Rev.} {\bf D66} (2002) 103511,
  [\href{http://arxiv.org/abs/astro-ph/0205436}{{\sf arXiv:astro-ph/0205436}}],
  [\href{http://dx.doi.org/10.1103/PhysRevD.66.103511}{{\sf
  doi:10.1103/PhysRevD.66.103511}}].

\bibitem{2017ARA&A..55..213S}
S.~{Sharma}, {\it {Markov Chain Monte Carlo Methods for Bayesian Data Analysis
  in Astronomy}},  {\sl Annual Review of Astronomy and Astrophysics} {\bf 55}
  (Aug., 2017) 213--259, [\href{http://arxiv.org/abs/1706.01629}{{\sf
  arXiv:1706.01629}}],
  [\href{http://dx.doi.org/10.1146/annurev-astro-082214-122339}{{\sf
  doi:10.1146/annurev-astro-082214-122339}}].

\bibitem{Planck:2013pxb}
{\bf Planck} Collaboration, P.~A.~R. Ade {\em et~al.}, {\it {Planck 2013
  results. XVI. Cosmological parameters}},  {\sl Astron. Astrophys.} {\bf 571}
  (2014) A16, [\href{http://arxiv.org/abs/1303.5076}{{\sf arXiv:1303.5076}}],
  [\href{http://dx.doi.org/10.1051/0004-6361/201321591}{{\sf
  doi:10.1051/0004-6361/201321591}}].

\bibitem{DES:2021wwk}
{\bf DES} Collaboration, T.~M.~C. Abbott {\em et~al.}, {\it {Dark Energy Survey
  Year 3 results: Cosmological constraints from galaxy clustering and weak
  lensing}},  {\sl Phys. Rev. D} {\bf 105} (2022), no.~2 023520,
  [\href{http://arxiv.org/abs/2105.13549}{{\sf arXiv:2105.13549}}],
  [\href{http://dx.doi.org/10.1103/PhysRevD.105.023520}{{\sf
  doi:10.1103/PhysRevD.105.023520}}].

\bibitem{PhysRevLett.95.011302}
S.~Wang, Z.~Haiman, W.~Hu, J.~Khoury, and M.~May, {\it Weighing neutrinos with
  galaxy cluster surveys},  {\sl Phys. Rev. Lett.} {\bf 95} (Jun, 2005) 011302,
  [\href{http://dx.doi.org/10.1103/PhysRevLett.95.011302}{{\sf
  doi:10.1103/PhysRevLett.95.011302}}].

\bibitem{Sagredo:2018ahx}
B.~Sagredo, S.~Nesseris, and D.~Sapone, {\it {Internal Robustness of Growth
  Rate data}},  {\sl Phys. Rev. D} {\bf 98} (2018), no.~8 083543,
  [\href{http://arxiv.org/abs/1806.10822}{{\sf arXiv:1806.10822}}],
  [\href{http://dx.doi.org/10.1103/PhysRevD.98.083543}{{\sf
  doi:10.1103/PhysRevD.98.083543}}].

\bibitem{ACT:2023dou}
{\bf ACT} Collaboration, F.~J. Qu {\em et~al.}, {\it {The Atacama Cosmology
  Telescope: A Measurement of the DR6 CMB Lensing Power Spectrum and its
  Implications for Structure Growth}},
  \href{http://arxiv.org/abs/2304.05202}{{\sf arXiv:2304.05202}}.

\bibitem{Knox:2019rjx}
L.~Knox and M.~Millea, {\it {Hubble constant hunter\textquoteright{}s guide}},
  {\sl Phys. Rev. D} {\bf 101} (2020), no.~4 043533,
  [\href{http://arxiv.org/abs/1908.03663}{{\sf arXiv:1908.03663}}],
  [\href{http://dx.doi.org/10.1103/PhysRevD.101.043533}{{\sf
  doi:10.1103/PhysRevD.101.043533}}].

\end{thebibliography}\endgroup

\end{document}